\documentclass[12pt]{article}
\usepackage{arydshln}
\usepackage{setspace}
\usepackage{hyperref}
\usepackage{amsmath}
\usepackage{color}
\usepackage {xcolor}
\usepackage{geometry}
\usepackage{listings}
\usepackage{booktabs}
\usepackage{mathrsfs}
\usepackage{multirow}
\newtheorem{thm}{Theorem}

\newtheorem{prop}{Proposition}

\usepackage{graphicx}
\usepackage{listings}
\usepackage{longtable}
\usepackage{mathrsfs}
\usepackage{amssymb}
\usepackage{bm}
\usepackage{caption}
\usepackage{authblk}
\usepackage{float}

\usepackage{hyperref}
\hypersetup{ colorlinks = true, linkcolor = blue, citecolor = blue}  
\usepackage{natbib}  
 \renewcommand{\citep}[1]{\citeauthor{#1}, \citeyear{#1}}  

\title{Model-Assisted Inference for {Covariate-Specific Treatment Effects} with High-dimensional Data}

\author[a]{Peng Wu}
\author[b]{Zhiqiang Tan\thanks{joint-first-author}}
\author[c]{Wenjie Hu}
\author[d]{Xiao-Hua Zhou\thanks{Corresponding author: azhou@math.pku.edu.cn}}
\affil[a]{Beijing International Center for Mathematical Research, \par Peking University,  Beijing 100871, China}

\affil[b]{Department of Statistics, Rutgers University, 110 Frelinghuysen Road, \par
 		Piscataway New Jersey 08854, U.S.A.}

\affil[c]{Department of Probability and Statistics,  
       Peking University,  Beijing 100871,  China }

\affil[d]{Department of Biostatistics, Beijing International Center for Mathematical Research \par 
  and National Engineering Laboratory of Big Data Analysis and Applied Technology, \par
   Peking University, Beijing 100871, China}

 \date{}

\begin{document}
\begin{spacing}{1.2}
  \maketitle

\begin{abstract}
	Covariate-specific treatment effects (CSTEs) represent heterogeneous treatment effects across subpopulations defined by certain selected covariates. 
 In this article, {we consider marginal structural models where CSTEs are linearly represented using a set of basis functions of the selected covariates.}
	We develop a new 
	 approach  in high-dimensional settings to obtain not only doubly robust point estimators of CSTEs,   but also model-assisted confidence intervals, which are valid when a propensity score model is correctly specified but an outcome regression model may be misspecified.  With a linear outcome model and subpopulations defined by discrete covariates, both point estimators and confidence intervals are doubly robust  for CSTEs.
	 In contrast,  confidence intervals from existing high-dimensional methods are valid only when both the propensity score and outcome models are correctly specified.  We establish asymptotic properties of the proposed point estimators and the associated confidence intervals.
	  We present  simulation studies and empirical applications which demonstrate the advantages of the proposed method compared with competing ones.
\end{abstract}

\section{Introduction}

When analyzing the causal effect of an intervention,  the average treatment effect (ATE) is often taken to be the estimand of interest for simplicity and interpretation. However,  researchers and policy makers can also be interested in  the effects of treatments (or policies)
		at various subpopulations levels
(\citep{Abrevaya-Hsu-Lieli2015};  \citep{Lee-Okui-Whang2017}; \citep{Chernozhukov-Luo2018}; \citep{Zimmert-Lechner2019}).   
 Specifically, let $Y$ be an outcome variable, $T$ be a treatment variable taking values in $\{0, 1\}$, and $Z$ be the covariates used to define subpopulations.  Define $(Y^0, Y^1)$ as the potential outcomes which would be observed under the treatment arms 0 and 1 respectively.  Of interest in this paper is the {covariate-specific treatment effect (CSTE)}  $\tau(z)$,
  defined by $E(Y^{1} - Y^{0} | Z=z)$ for possible values $z$ of $Z$.
   For example,  in our empirical application, we study the effects of maternal smoking on  infant birth weights in different subpopulations  defined by mother's age.   In clinical settings,
    {CSTEs
  are useful  in precision medicine for the discovery of optimal treatment regimes that can be tailored to individual's characteristics (\citep{Chakraborty-Moodie2013}) }.

For observational studies, a large set of covariates are often included, possibly with nonlinear and interaction terms, in statistical analysis
to reduce confounding bias and enhance the credibility of causal inference.
  Thus, we introduce auxiliary covariates $V$, allowing $V$ to be high-dimensional, and posit that the unconfoundedness holds conditionally on all covariates $X \equiv (Z, V)$ to obtain the identification of {CSTEs}.

  The {CSTE} $\tau(z)$  is in general  different from  $\tau(x) \equiv E(Y^{1} - Y^{0} | X=x)$, the conditional treatment effect given the full covariates. 
     Being conditional on a low-dimensional covariate,
  $\tau(z)$ is easier to interpret and communicate in practice.  Moreover, estimation of $\tau(z)$ can be more manageable and less affected by modeling assumptions in statistical analysis.
      It is known to be difficult to obtain asymptotic normality and valid confidence intervals for
          $\tau(x)$ due to the high dimensionality of $X$, unless some restrictive assumptions 
          are imposed (\citep{Tian-etal2014}; \citep{Zhao-Small-Ertefaie2018}; \citep{Dukes-Vansteelandt2020}; \citep{Guo-Zhou-Ma2021}).
 
 There has been increasing interest in estimating {CSTEs} in recent years.
	\citet{Abrevaya-Hsu-Lieli2015} derived an inverse probability weighting (IPW) estimator of $\tau(z)$ using kernel smoothing with continuous $Z$,
       \citet{Lee-Okui-Whang2017} proposed an AIPW (augmented IPW) estimator based on kernel smoothing, and \cite{Lechner2019}  proposed algorithms to  construct causal random  forests.  These three approaches estimate $\tau(z)$ in low-dimensional settings. 
        \citet{Fan-Hsu-Lieli-Zhang2019} and \citet{Zimmert-Lechner2019}  extended the method of  \citet{Lee-Okui-Whang2017} to high-dimensional settings.  The authors adopted machine learning algorithms to mitigate  model specification for nuisance parameters (PS and OR models), and used sample splitting (or cross-fitting) technique to reduce the impact of nuisance parameters estimation on the resulting estimator of $\tau(z)$.
       A limitation of these existing high-dimensional methods 
       is that the confidence intervals are shown to be valid 
       when both PS and OR models are correctly specified. Further discussion is provided in Section  \ref{sec-existing-estimators}.

 Our proposed method is motivated by \citet{Tan-Annals2020}, 
 where a novel method is developed  to obtain not only doubly robust point estimators for ATEs in high-dimensional settings,
   but also model-assisted confidence intervals, which are valid when a propensity score (PS) model is correctly specified but  an outcome regression (OR) model may be misspecified. With a linear OR model, the confidence intervals are also doubly robust.
The method of \citet{Tan-Annals2020} is first proposed to estimate ATEs and  average treatment effects on treated (ATTs), and
recently extended to estimate local average treatment effects (LATEs)  in high-dimensional settings  (\citep{Sun-Tan2020}). In this article, we further extend the method to tackle estimation of {CSTEs}, which differ from the former quantities in that $\tau(z)$ is a function of the covariate value $z$.

  {To handle CSTEs defined by continuous or discrete covariates $Z$  or a combination of them, we consider
  marginal structural models, where CSTEs are linearly represented using a set of basis functions in $z$ (\citep{Robins-1999}; \citep{Tan-2010}).
 For discrete covariates $Z$, these models are unrestrictive when saturated basis functions are used.
 For continuous covariates $Z$, these models can be used to provide sufficiently accurate approximations with flexible basis functions.}
 We propose both doubly robust point estimators and model-assisted confidence intervals for
 {CSTEs} in high-dimensional settings.
   Remarkably,  the model-assisted confidence intervals can be derived  by a careful specification of regressors in fitting the OR model.
 	In addition,  with a linear OR model and discrete $Z$, we  obtain doubly robust confidence intervals
	by adding a full set of interactions between $Z$ and $V$  into the regressors when fitting the PS model.
      To the best of our knowledge, there is no method for estimating {CSTEs} that possesses these  desired properties including model-assisted and doubly robust confidence intervals. 

The rest of the article is structured as follows. In Section \ref{sec-background}, we state the setup of   problem interested and discuss some existing methods. Section \ref{sec-methods} presents  our estimation procedures in details. 
Section \ref{sec-asymptotic} shows the asymptotic results and elucidates why the proposed methods work.  In Section \ref{sec-simu}, extensive simulations are conducted to evaluate the finite sample performance of the proposed methods.  Section \ref{sec-application} illustrates our methods with an empirical example.  A brief discussion is presented in Section \ref{sec-discussion}.


\section{Background}  \label{sec-background}

\subsection{Setup}

Suppose that $\{(Y_{i}, T_{i}, X_{i}): i= 1, ..., n \}$ is an independent and identically distributed sample of $n$ observations, where  $Y$ is an outcome variable, $T$ is a treatment variable taking values in $\{0, 1\}$, and $X = (V^{T}, Z^{T})^{T}$ is a vector of measured covariates, where $Z$ is the covariates used to define subpopulations, $V$ is auxiliary covariates. In the potential outcomes framework (\citep{Rubin1974}; \citep{Neyman-1990}),  let
   $(Y^0, Y^1)$ be the potential outcomes under the treatment arms 0 and 1 respectively. By the consistency assumption, the observed outcome is $Y = (1-T)Y^{0} + T Y^{1}$. The causal parameter of interest is the {CSTE} defined by $\tau(z) = E(Y^{1} - Y^{0} | Z = z) = \mu^{1}(z) - \mu^{0}(z)$ with $\mu^{t}(z) = E(Y^{t}  | Z = z)$ for $t=0, 1$.
For identification of $(\mu^{0}(z), \mu^{1}(z))$, two common assumptions are imposed throughout:
	\begin{itemize}
	\item[(i)] Unconfoundedness: $T \perp Y^{0} | X$ and $T \perp Y^{1} | X$ (\citep{Rubin1976}).
	\item[(ii)] Overlap: $0 < \pi^{*}(x) < 1$ for all $x$, where $\pi^{*}(x) = P(T=1| X= x)$ is called propensity score (\citep{Rosenbaum-Rubin1983}).
	\end{itemize}	
Under these assumptions, letting $m_{t}^{*}(X)  =  E(Y| T = t, X)$, we have
	\begin{align}   \label{eq1}
			\mu^{1}(z)
			 ={}&  E[   m_{1}^{*}(X)    |  Z = z     ]  = E \Big [  \frac{TY}{\pi^{*}(X)}   |  Z = z \Big ]  \notag   \\
			={}& E \Big [   \frac{   TY  }{ \pi^{*}(X) } -( \frac{T}{\pi^{*}(X)} -1 ) m_{1}^{*}(X)     \big |   Z = z               \Big ].			
	\end{align}
These identification results follow from 
  direct applications of  the law of iterated expectations. 
  Similar equations  can be derived  for $\mu^{0}(z)$ and $\tau(z)$. Then, $(\mu^{0}(z), \mu^{1}(z))$ and $\tau(z)$ can be estimated by imposing additional modeling assumptions on the outcome regression (OR) function $m_{t}^{*}(X)$ or the propensity score (PS) $\pi^{*}(X)$.     We mainly discuss estimation of $\mu^{1}(z)$ and defer the discussion about   $\mu^{0}(z)$ and $\tau(z)$ to Section \ref{sec-3.4}.

\subsection{Existing estimators}   \label{sec-existing-estimators}

Consider a conditional mean model for OR in the treated group,
			\begin{equation}  \label{eq2}
						E(Y| T=1, X) = m_{1}(X; \alpha_{1}) = \psi\{ \alpha^{T}_{1} g(X)   \},
			\end{equation}
where $\psi(\cdot)$ is an inverse link function, $g(X) = \{1, g_{1}(X), ..., g_{q}(X)  \}^{T}$ is a vector of known functions such as $g(X) = (1, X^{T})^{T}$.  Throughout, the superscript $^{T}$ denotes a transpose, not the treatment variable $T$.
 Alternatively, consider a PS model
		\begin{equation}  \label{eq3}
					P( T=1 | X) = \pi(X; \gamma) = \Pi \{ \gamma^{T} f(X)   \},
			\end{equation}
where  $\Pi(\cdot)$ is an inverse link function, $f(X) = \{1, f_{1}(X), ..., f_{p}(X)  \}^{T}$ is a vector of known functions. 
 	 For concreteness, assume that model (\ref{eq3}) is logistic regression with $\pi(X; \gamma) = [ 1 + \exp\{ -\gamma^{T} f(X) \}  ]^{-1}$.

For OR model (\ref{eq2}), the average negative log-(quasi-)likelihood function can be written as
 		\begin{equation}     \label{eq4}
				L_{ML}(\alpha_{1})  =  \tilde E \big (   T   [  - Y \alpha^{T}_{1} g(X)  +  \Psi \{  \alpha^{T}_{1} g(X)  \}    \big   )
		\end{equation}
 where $\Psi(u) = \int_{0}^{u} \psi(u')\text{d}u'$ and $\tilde E(\cdot) $ denotes the sample average throughout.    In high-dimensional settings,  a lasso penalized maximum likelihood estimator  $\hat \alpha_{1, RML}$  can be  defined as a minimizer of  $L_{RML}(\alpha_{1}) = L_{ML}(\alpha_{1}) + \lambda || (\alpha_{1})_{1:q} ||_{1}$,
    where $||\cdot||_{1}$ denotes the $L_{1}$ norm, $(\alpha_{1})_{1:q}$ is $\alpha_{1}$ excluding the intercept,  $\lambda \geq 0$ is a tuning parameter.  Let $\hat m_{1, RML}(X) = m_{1}(X; \hat \alpha_{1, RML})$.
    Then an outcome-regression based estimator of $\mu^{1}(z)$ can be derived by  regressing $\hat m_{1, RML}(X)$  on $Z$. To be specific, for a continuous covariate $Z$,  the local constant regression
   (\citep{Li-Racine2007})  leads to
 		$$    \hat \mu^{1}_{OR}(z; \hat m_{1, RML})  =   \frac{ \sum_{i=1}^{n} K( \frac{Z_{i} - z}{h})  \hat m_{1, RML}(X_{i}) }   {  \sum_{i=1}^{n} K( \frac{Z_{i} - z}{h})  },    $$
where $K(u)$ is a kernel function and $h$ is a bandwidth.

Alternatively, for PS model (\ref{eq3}),  the negative log-likelihood function is
			\begin{equation}  \label{eq5}
				L_{ML}(\gamma)  =  \tilde E \big [   - T \gamma^{T} f(X)  + \log\{ 1 + e^{\gamma^{T} f(X)}  \}    \big   ].
		\end{equation}
Define  $\hat \gamma_{RML}$ as a lasso penalized maximum likelihood estimator of $\gamma$ which is a minimizer of $L_{ML}(\gamma)   +  \lambda ||\gamma_{1:p}||_{1}$  to handle high-dimensional data, where $\gamma_{1:p}$ is $\gamma$ excluding the intercept. Denote $\hat \pi_{RML}(X) = \pi(X; \hat \gamma_{RML})$. Then in the spirit of  \citet{Abrevaya-Hsu-Lieli2015}, for a continuous covariate $Z$, an inverse probability weighted (IPW) estimator of $\mu^{1}(z)$ can be obtained by conducting  local constant regression  $T Y /  \hat \pi_{RML}(X)$ on $Z$:
	      \[   \hat \mu^{1}_{IPW}(z; \hat \pi_{RML}) =   \frac{ \sum_{i=1}^{n} K( \frac{Z_{i} - z}{h}) T_{i} Y_{i} / \hat \pi_{RML}(X_{i})  }   {  \sum_{i=1}^{n} K( \frac{Z_{i} - z}{h})  }.                   \]

Consistency of the estimator $ \hat \mu^{1}_{OR}(z; \hat m_{1, RML})$ or  $\hat \mu^{1}_{IPW}(z; \hat \pi_{RML})$
   relies on the correct specification of  OR model (\ref{eq2}) or PS model (\ref{eq3}),  respectively.   Doubly robust estimators can be constructed in the augmented IPW (AIPW) form  by combining OR and PS models (\citep{Robins-Rotnitzky-Zhao1994}; \citep{Kang-Schafer2007}; \citeauthor{Tan2007} \citeyear{Tan2007,Tan2010}).  Let
		\begin{equation} \label{eq6} 	\varphi(Y_{i}, T_{i}, X_{i};  \hat m_{1, RML}, \hat \pi_{RML}) =  \frac{   T_{i} Y_{i}  }{ \hat \pi_{RML}(X_{i}) } -( \frac{T_{i}}{\hat \pi_{RML}(X_{i})} -1 ) \hat m_{1, RML}(X_{i}).
        \end{equation}
Equation  (\ref{eq1}) implies that the doubly robust  AIPW estimator of $\mu^{1}(z)$ can be obtained via regressing $\varphi(Y_{i}, T_{i}, X_{i};  \hat m_{1, RML}, \hat \pi_{RML})$ on $Z$.  See \citet{Lee-Okui-Whang2017} in low-dimensional settings, {and}
\citet{Fan-Hsu-Lieli-Zhang2019} and \citet{Zimmert-Lechner2019} in high-dimensional settings.    For instance, a local constant estimator of $\mu^{1}(z)$ is
		    \[  \hat \mu^{1}(z; \hat m_{1, RML}, \hat \pi_{RML}) =   \frac{ \sum_{i=1}^{n} K( \frac{Z_{i} - z}{h})   \varphi(Y_{i}, T_{i}, X_{i};  \hat m_{1, RML}, \hat \pi_{RML})  }   {  \sum_{i=1}^{n} K( \frac{Z_{i} - z}{h})  }.                              \]
 {These authors also adopted machine learning algorithms to fit flexible PS and OR models, and used sample splitting technique to reduce the impact of parameter estimation in PS and OR models
 on the resulting estimator of $\mu^{1}(z)$.}

    Compared with $\hat \mu^{1}_{OR}(z; \hat m_{1, RML})$ and $\hat \mu^{1}_{IPW}(z; \hat \pi_{RML})$, {in addition to being doubly robust point estimator}, there is potentially a further advantage of $ \hat \mu^{1}(z; \hat m_{1, RML}, \hat \pi_{RML})$ in high-dimensional settings.    Since both $\hat m_{1, RML}$ and $\hat \pi_{RML}$ usually converge at a slower rate than $O_{p}( 1/\sqrt{nh})$ for high-dimensional $X$,  the resulting convergence rates for  $\hat \mu^{1}_{OR}(z; \hat m_{1, RML})$ and $\hat \mu^{1}_{IPW}(z; \hat \pi_{RML})$ will be slower than  $O_{p}(1/\sqrt{nh})$.
     According to \citet{Zimmert-Lechner2019}, if both models (\ref{eq2}) and (\ref{eq3}) are correctly specified {or with negligible biases,  then under suitable conditions, }  $\hat \mu^{1}(z; \hat m_{1, RML}, \hat \pi_{RML})$ converges to $\mu^{1}(z)$ at rate $O_{p}(1/\sqrt{nh})$ and admits an asymptotic expansion
		\begin{equation} \label{eq7}   \hat \mu^{1}(z; \hat m_{1, RML}, \hat \pi_{RML}) =      \frac{ \sum_{i=1}^{n} K( \frac{Z_{i} - z}{h})   \varphi(Y_{i}, T_{i}, X_{i};  m_{1}^{*}, \pi^{*}) }   {  \sum_{i=1}^{n} K( \frac{Z_{i} - z}{h})  }     + R_{n}(z),    \end{equation}
 where $R_{n}(z) =  o_{p}(  1/\sqrt{nh} )$. 
 However, when only one of the model (\ref{eq2}) or  (\ref{eq3}) is correctly specified, the asymptotic {expansion} (\ref{eq7}) or
 the associated confidence interval for $\mu^1(z)$ does not in general hold. 
 


\section{Methods}   \label{sec-methods}
We develop new methods to obtain both doubly robust point estimators and model-assisted confidence intervals for $(\mu^1(z), \mu^0(z))$ and $\tau(z)$,
based on marginal structural models (\citep{Robins-1999}; \citep{Tan-2010}).  

{We first discuss estimation of $\mu^1(z)$. 
Let $\Phi(z) = (\phi_{1}(z), ..., \phi_{K}(z))^{T}$ be a vector of basis functions excluding the constant. 
Consider a marginal structural model where $\mu^1(z)$ is linearly represented as 
					\begin{equation} \label{eq16}
							\mu^{1}(z)  =  \beta_{0}^{*} +  \beta_{1}^{*T} \Phi(z) 
					\end{equation}  }
where $\beta^{*} = (\beta_{0}^{*}, \beta_{1}^{*T})^{T}$ is a vector of parameters.
Different choices of $\Phi(z)$ can be used, to accommodate different data types of the covariates $Z$ as follows.
\begin{itemize}
\item[(a)]    $Z$ is a binary variable. {Let $\Phi(z)= z$. Then model (\ref{eq16}) is saturated.}

\item[(b)]  $Z$ is a categorical variable taking multiple values. For example, suppose that $Z$ is a trichotomous variable encoded as two dummy variables $(Z_{1}, Z_{2})$.
  {Let $\Phi(z)= ( z_{1}, z_{2})^{T}$. Then model (\ref{eq16}) saturated}.

\item[(c)]  $Z$ consists of multiple binary variables. Suppose that $Z = (Z_{1}, Z_{2})$, where $Z_{1}$ and $Z_{2}$ are two binary variables.  
{Let $\Phi(z)= (z_{1}, z_{2}, z_{1} z_{2})^{T}$. Then model (\ref{eq16}) saturated.}
		 	Importantly,  when $Z$ consists of multiple discrete variables,  it also can be encoded as  multiple binary variables.

\item[(d)]  $Z$ is a continuous variable. Then {$\Phi(z)$ can be specified using 
spline basis (\citep{Schumaker2007}) and Fourier basis (\citep{Ramsay-Silverman2005}) similarly as in nonparametric estimation of a regression curve.}

\item[(e)]  $Z$ is a combination of discrete and continuous variables, for example, $Z = (Z_{1}, Z_{2})$, where $Z_{1}$ is a binary variable  and $Z_{2}$ is a continuous variable.
Then we can set $\Phi(z) = ( z_{1},  B^{T}(z_{2}),  z_{1} B^{T}(z_{2}) )^{T}$, where $B(z_{2})$ consists of basis functions of $Z_{2}$.
\end{itemize}

{
Model (\ref{eq16}) can be made to be saturated by a proper choice of $\Phi(z)$ for a discrete $Z$.
But for a continuous $Z$, model (\ref{eq16}) with a fixed set of basis functions may not hold exactly,
i.e., $\mu^{1}(z)$ may not fall in the class $\{\beta_{0} + \beta_{1}^{T} \Phi(z):   (\beta_{0}, \beta_{1}) \in \mathbb{R}^{K+1}  \}$.
In this case, model (\ref{eq16}) can be interpreted such that
 $ \beta_{0}^{*} +  \beta_{1}^{*T} \Phi(z)$ gives the best linear approximation of $\mu^{1}(z)$  using basis functions (1, $\Phi(z)$), where 
\begin{align}
  (\beta_{0}^{*}, \beta_{1}^{*}) 
  =    \arg \min_{\beta_{0}, \beta_{1}} E \big ( \mu^{1}(Z) -  \beta_{0} - \beta_{1}^{T} \Phi(Z) \big )^{2}.  \label{eq:beta*}
\end{align}
As shown in our simulation study (Section 5), 
the proposed method is expected to perform well when model (\ref{eq16}) provides a sufficiently accurate approximation with a flexible choice of basis functions in $\Phi(z)$. }

\subsection{Regularized calibrated estimation}

{To focus on main ideas, consider the generalized linear model (\ref{eq2}) and the logistic propensity score model
			\begin{equation}  \label{eq8}
				P(T=1| X) = \pi(X; \gamma) = [  1 + \exp \{ -\gamma^{T} f(X) \}  ]^{-1}.
			\end{equation}}
Instead of using regularized likelihood estimation in Section \ref{sec-existing-estimators},  we adopt
the regularized calibrated estimator of $\gamma$  and regularized weighted  likelihood estimator of $\alpha_{1}$ (\citeauthor{Tan-BKA2020}, \citeyear{Tan-Annals2020, Tan-BKA2020}).
For PS model (\ref{eq8}), the  regularized calibrated estimator $\hat \gamma_{RCAL}$ is defined as a minimizer of the lasso penalized objective function,
		\begin{equation} \label{eq9}     L_{RCAL}(\gamma) =  L_{CAL}(\gamma)   + \lambda  || \gamma_{1:p} ||_{1},                    \end{equation}
 where $L_{CAL}(\gamma)$  is the calibration loss
 		\begin{equation}    L_{CAL}(\gamma) =  \tilde E\{   T e^{-\gamma^{T} f(X)} + (1- T) \gamma^{T} f(X)      \}.    \end{equation}
 For OR model (\ref{eq2}),  the regularized weighted likelihood estimator  $\hat \alpha_{1, RWL}$ is defined as a minimizer of the lasso penalized objective function
	\begin{equation}    \label{eq11}
			L_{RWL}(\alpha_{1}; \hat \gamma_{RCAL}) = L_{WL}(\alpha_{1}; \hat \gamma_{RCAL}) + \lambda ||  (\alpha_{1})_{1:q} ||_{1},
	\end{equation}  
with the weighted likelihood loss
	\begin{equation}  \label{eq12}
		L_{WL}(\alpha_{1}; \hat \gamma_{RCAL}) = \tilde E \big (    T w(X; \hat \gamma_{RCAL})[ - Y \alpha_{1}^{T} f(X) + \Psi\{ \alpha_{1}^{T} f(X) \} ]  \big  ),	
	\end{equation}	
where   $\Psi(u) = \int_{0}^{u} \psi(u')\text{d}u'$, $w(X; \gamma) = \{1 - \pi(X; \gamma)  \} / \pi(X; \gamma) = e^{-\gamma^{T} f(X)}$.  Let $\hat \pi_{RCAL}(X) = \pi(X; \hat \gamma_{RCAL})$ be the fitted PS function and  $\hat m_{1, RWL}(X) = m_{1}(X; \hat \alpha_{1, RWL})$ be the fitted OR function. {As indicated by (\ref{eq12}), $\hat m_{1, RWL}(X)$ depends on  $\hat \pi_{RCAL}(X)$,
 in contrast with the recent papers of \cite{Fan-Hsu-Lieli-Zhang2019} and \cite{Zimmert-Lechner2019},  where the propensity score and outcome regression functions are estimated separately.}

   Before proceeding to the main ideas,
 we present several interesting properties algebraically associated with $\hat \pi_{RCAL}(X)$ and $\hat m_{1, RWL}(X)$,  part of which are also used in proving our results later.
  First,  by the Karush-Kuhn-Tucker (KKT) condition for minimizing (\ref{eq9}), the fitted propensity score  $\hat \pi_{RCAL}(X)$ satisfies
	\begin{align}
		&		\frac 1 n \sum_{i=1}^{n} \frac{ T_{i}}{  \hat \pi_{RCAL}(X_{i}) }   =  1,  \label{eq13} \\
		&      \frac 1 n  \biggl | \sum_{i=1}^{n} \frac{ T_{i}  f_{j}(X_{i})  }{  \hat \pi_{RCAL}(X_{i})    }  - \sum_{i=1}^{n} f_{j}(X_{i})  \biggr |  \leq  \lambda,  \quad j = 1, ..., p. 	   \label{eq14}
	\end{align}
where equality holds in (\ref{eq14}) for any $j$ such that the $j$-th element of $\hat \gamma_{RCAL}$ is nonzero.   Equation (\ref{eq13}) shows that the sum of inverse probability weights $T /  \hat \pi_{RCAL}(X) $ equals to sample size $n$, whereas  equation (\ref{eq14}) implies that the weighted average of each covariate $f_{j}(X_{i})$ over the treated group may differ from the overall average of $f_{j}(X_{i})$ by no more than $\lambda$. In addition,  \cite{Tan-BKA2020} showed that, with possible model misspecification,  calibrated estimation is better than maximum likelihood estimation in terms of controlling mean squared errors of inverse probability weighted estimators.

Second,  by the KKT condition for minimizing (\ref{eq11}), the  fitted treatment regression function $\hat m_{1, RWL}(X)$ satisfies
       	\begin{equation}
				n^{-1} \sum_{i=1}^{n} T_{i}   w(X; \hat \gamma_{RCAL})  \big \{ Y_{i}  - \hat m_{1, RWL}(X_{i}) \big \}   =  0.  \label{eq15}
	\end{equation}
As a consequence of equation (\ref{eq15}),  the augmented IPW estimator for $E(Y^{1})$, defined as $\hat E_{RCAL}(Y^{1}) = \tilde E\{  \varphi(Y, T, X;  \hat m_{1, RWL} , \hat \pi_{RCAL} ) \}$, can be reformulated as
	\[   \tilde E \biggl [    \hat m_{1, RWL}(X)  +  \frac{T}{ \hat \pi_{RCAL}(X) }  \big\{ Y - \hat m_{1, RWL}(X) \big \}    \biggr ]    =  \tilde E \{    T Y  + (1- T)\hat  m_{1, RWL}(X)       \},                \]
 which implies that $\hat E_{RCAL}(Y^{1})$ always fall within the range of  the observed outcomes $\{Y_{i}: T_{i} = 1, i = 1, ..., n \}$ and the predicted values $\{ \hat m_{1, RWL}(X_{i}): T_{i} = 0, i = 1, ...,  n  \}$.


\subsection{Model-assisted confidence intervals of {$\mu^{1}(z)$} } \label{sec-model-assisted}
 For ease of exposition hereafter,  we let $\hat \gamma = \hat \gamma_{RCAL}$, $\hat \alpha_{1} = \hat \alpha_{1, RWL}$,
 $\hat \pi = \hat \pi_{RCAL}(X)$, $\hat m_{1} = \hat m_{1, RWL}(X)$,  $\hat \varphi =  \varphi(Y, T, X;  \hat m_{1, RWL} , \hat \pi_{RCAL} )$,  $\varphi^{*}  =\varphi(Y, T, X;   m^{*}_{1},  \pi^{*}) $, and
   {$\Phi^\dag(z) = (1, \Phi(z)^{T})^{T}$}.

By {the identity (\ref{eq1}) for $\mu^1(z)$ and the expression (\ref{eq:beta*}) for $(\beta_0^*, \beta_1^*)$},
it seems natural to define an estimator of $\beta^{*}$ directly as
\[   \hat \beta =  (\hat \beta_{0}, \hat \beta_{1}^{T} )^{T} =  \tilde E^{-1} \{  \Phi^\dag(Z)  \Phi^\dag(Z)^{T}  \} \tilde E\{  \Phi^\dag(Z) \hat \varphi  \}.                   \]
The corresponding estimator of {$\mu^{1}(z)$} is
	\begin{equation}   \label{eq17}
				\hat \mu^{1}(z; \hat m_{1}, \hat \pi) =  \hat \beta^{T} \Phi^\dag(z) ,
	\end{equation}
 The estimator $\hat \mu^{1}(z; \hat m_{1}, \hat \pi)$
 is easily shown to be a doubly robust point estimator of {$\mu^{1}(z)$}.
 Remarkably, model-assisted confidence intervals for {$\mu^1(z)$} can be derived
 by a careful specification of  $g(X)$ in fitting  OR model (\ref{eq2}).

 Define $f(X) \otimes \Phi(Z)$ as the vector of all the interactions between $f(X)$ and $\Phi(Z)$.  To obtain  model-assisted confidence intervals, {we set
		\begin{equation}    \label{eq18}   g(X) = (f(X)^{T}, (f(X) \otimes \Phi(Z))^{T} )^{T}.   \end{equation}}
There may be same functions repeated in $g(X)$. In that case,  we let $g(X)$ be the vector
$(f(X)^{T}, (f(X) \otimes \Phi(Z))^{T} )^{T}$ after excluding the duplicated elements.
To put it more clearly, we let   $f(X) = (1, V^{T}, \Phi(Z)^{T})^{T}$  and present the specific form of $g(X)$ for the first four data types of $Z$ after equation (\ref{eq16}):
	\begin{itemize}
	\item $Z$ is a binary variable,  $f(X) = (1, V^{T}, Z)^{T}$, $g(X) = (1, V^{T}, Z, V^{T} Z)^{T}$.
	\item  
	$Z$ is a trichotomous variable encoded as two dummy variables $(Z_{1}, Z_{2})$,
		$f(X) = (1, V^{T}, Z_{1}, Z_{2})^{T}$, $g(X)= ( 1, V^{T}, Z_{1},  Z_{2},  V^{T}Z_{1},  V^{T} Z_{2}  )^{T}$.
	\item $Z$ consists of two binary variables $Z_{1}$ and $Z_{2}$. $f(X) = (1, V^{T}, Z_{1}, Z_{2},  Z_{1}Z_{2})^{T}$, $g(X)=	 (1, V^{T}, Z_{1}, Z_{2}, Z_{1}Z_{2},  V^{T} Z_{1},  V^{T} Z_{2},  V^{T} Z_{1}Z_{2} )^{T}$.
	\item $Z$ is a continuous variable,  $f(X) = (1, V^{T}, \Phi(Z)^{T})^{T}$, $g(X) = ( 1, V^{T},
	\Phi(Z)^{T}, (V \otimes \Phi(Z))^{T} ,  (\Phi(Z) \otimes \Phi(Z) )^{T} )^{T}$.
	\end{itemize}
In general, the choice of $f(X)$ can be flexible.  
	For instance,  it is possible to include full interactions between $V$ and $\Phi(Z)$ in $f(X)$, namely,  $f(X) = (1, V^{T}, \Phi(Z)^{T},  (V \otimes \Phi(Z))^{T} )^{T}$.
	Interestingly, this choice of $f(X)$ can be applied to construct doubly robust confidence intervals  for $\mu^1(z)$ with discrete $Z$, as shown in Section \ref{doubly-robust}.
 In addition, it is possible to include more  covariates, such as nonlinear terms of $V$, in $f(X)$. These additional terms are easily accommodated under sparsity conditions.

We  provide a high-dimensional analysis of the estimator $\hat \mu^{1}(z; \hat m_{1}, \hat \pi)$ in (\ref{eq17}), allowing for possible model misspecification. Define $\bar \gamma $  as a minimizer of the expected calibration loss	
		 $E\{ L_{CAL}(\gamma)  \} =  E\{  T e^{-\gamma^{T} f(X)}  + (1- T) \gamma^{T} f(X)       \}$
and $\bar \alpha_{1}$ as a minimizer of
		\[   E\{ L_{WL}(\alpha_{1}; \bar \gamma)   \} = E\big [ T  w(X; \bar \gamma)   [   - Y \alpha_{1}^{T} f(X) + \Psi\{ \alpha_{1}^{T} f(X) \}    ]  \big ].      \]
Let  $\bar \pi =  \pi(X; \bar \gamma)$, $\bar  m_{1} = m(X; \bar \alpha_{1})$ and $\bar \varphi =\varphi(Y, T, X;   \bar m_{1},  \bar \pi)$.   	
If model  (\ref{eq8}) is correctly specified, then $\bar \pi = \pi^{*}$; otherwise, $\bar \pi$ may different from $\pi^{*}$.  Likewise, if model (\ref{eq2}) is correctly specified,  then $\bar m_{1} = m_{1}^{*}$; $\bar m_{1}  \neq m_{1}^{*}$ otherwise.  Let
	 \[   \bar \beta \equiv ( \bar \beta_{0},  \bar \beta_{1}^{T})^{T}  =  \tilde E^{-1} \{  \Phi^\dag(Z)  \Phi^\dag(Z)^{T}  \} \tilde E\{  \Phi^\dag(Z)  \bar \varphi  \},                   \]
	and 	$\hat \mu^{1}(z; \bar m_{1}, \bar \pi) =  \bar \beta^{T} \Phi^\dag(z).$
 Our main result shows that under suitable conditions,
	\begin{equation}  \label{eq19}
	\hat \mu^{1}(z; \hat m_{1}, \hat \pi) = \hat \mu^{1}(z; \bar m_{1}, \bar \pi) +   R_{n}(z),
	\end{equation}
with {$ |R_{n}(z)| = o_{p}(n^{-1/2})$} for both discrete $Z$ and continuous $Z$.

Suppose that the lasso tuning parameters are specified as $\lambda = A_{0} \lambda_{0}$ for $\hat \gamma$ and $\lambda = A_{1} \lambda_{1}$ for $\hat \alpha_{1}$, where
  $A_{0}$ and $A_{1}$ are two sufficiently large positive constants,  the tuning parameters $(\lambda_{0}, \lambda_{1})$ are set as   {
	 			$\lambda_{0} =  \sqrt{ \log \{ (1+ p) /  \epsilon   \}  / n  },$
				$  \lambda_{1} =\sqrt{ \log \{ (1+ q) /  \epsilon   \}  / n  } ~ (\geq \lambda_{0}),$	 }
 where $0 < \epsilon < 1$ is a tail probability for the error bound. For example, $\lambda_{0} = \sqrt{ 2 \log (1 + p) / n}$ by taking $\epsilon = 1 / (1 + p)$.
   For a vector $b = (b_{0}, b_{1}, ..., b_{p})^{T}$, denote $S_{b} = \{0\} \cup \{j: b_{j} \neq 0, j = 1, ..., p  \}$ and the size of the set $S_{b}$ as $|S_{b}|$.
 { The following Propositions \ref{prop1} can  be deduced from Theorem 3 directly. }

\begin{prop}[Model-assisted confidence intervals] \label{prop1}  Suppose that Assumptions 1--2 hold as in Section \ref{sec-theory}, $g(X)$ is chosen as in (\ref{eq18}), and
{$( |S_{\bar \gamma}| + |S_{\bar \alpha_1}| ) \log(q)  = o(n^{1/2})$}. If PS model (\ref{eq8}) is correctly specified, then asymptotic expansion (\ref{eq19}) is valid.     Furthermore, for any given $z_{0}$,  the following results hold for both discrete $Z$ and continuous $Z$:

 (i) $n^{1/2} \{ \hat \mu^{1}(z_{0}; \hat m_{1}, \hat \pi)  -  \mu^{1}(z_{0})   \} \xrightarrow{D}  N(0, V(z_{0}) )$, where
		$$ V(z_{0}) =  \text{var} \Big \{  \Phi^\dag(z_{0})^{T}     E^{-1} \big [ \Phi^\dag(Z) \Phi^\dag(Z)^{T}    \big ]       \Phi^\dag(Z)  \varphi(Y, T, X; \bar m_{1}, \bar \pi )       \Big \}.   $$

   (ii) a consistent estimator of $V(z_{0})$ is  	
      $$ \hat V(z_{0})  =    \Phi^\dag(z_{0})^{T}  M^{-1}  \hat  G   M^{-1}   \Phi^\dag(z_{0})  / n,    $$
where $M =  \tilde E \{  \Phi^\dag(Z)  \Phi^\dag(Z)^{T}  \}$ and
		\[ \hat G  =  n^{-1}  \sum_{i=1}^{n}  \Big \{  \Phi^\dag(Z_{i})    \Phi^\dag(Z_{i})^{T}   [   \varphi( Y_{i}, T_{i}, X_{i}; \hat m_{1}, \hat \pi)  -   \hat \beta^{T}  \Phi^\dag(Z_{i})    ]^{2}     \Big \}.        \]	

   (iii) an asymptotic $(1- c)$ confidence interval for $\mu^{1}(z_{0})$ is $\hat \mu(z_{0}; \hat m_{1}, \hat \pi_{1})
   \pm z_{c/2} \sqrt{ \hat V(z_{0}) / n } $, where $z_{c/2}$ is the $(1- c/2)$ quantile of $N(0, 1)$. That is, a model-assisted confidence interval for $\mu^{1}(z_{0})$ is obtained.

\end{prop}

{For simplicity, the preceding result is stated while assuming that model (\ref{eq16}) is correct.
If model (\ref{eq16}) does not hold exactly, then the confidence interval $\hat \mu(z_{0}; \hat m_{1}, \hat \pi_{1})  \pm z_{c/2} \sqrt{ \hat V(z_{0}) / n }$
remains valid when evaluated against the approximate value $\tilde \mu^1(z)= \beta_{0}^{*} +  \beta_{1}^{*T} \Phi(z)$ for $(\beta_0^*, \beta_1^*)$ defined in (\ref{eq:beta*}).}
In Section \ref{sec-simu}, our simulation study shows that the approximate confidence intervals perform very well.
  
   Why can the model-assisted confidence intervals be obtained by a careful specification of $g(X)$? We provide a theoretical analysis of the estimator $\hat \mu^{1}(z; \hat m_{1}, \hat \pi)$  in Section \ref{sec-asymptotic}, where Section \ref{sec-heuristic} gives a heuristic discussion.

\subsection{Doubly robust confidence intervals of $\mu^{1}(z)$ for discrete $Z$} 	  \label{doubly-robust}
We derive doubly robust confidence intervals {for $\mu^1(z)$ with discrete $Z$} when a linear OR model is used. Consider the linear OR model
\begin{equation}    \label{eq20}
			 E(Y | T= 1, X)  = m_{1}(X; \alpha_{1})  =  \alpha^{T}_{1} g(X)
	\end{equation}
and the PS model (\ref{eq8}). Remarkably, doubly robust confidence intervals for $\mu^1(z)$ can be obtained merely by  including full interactions between $V$ and $\Phi(Z)$ in $f(X)$, that is, setting
	\begin{equation}    \label{eq21}   f(X) = (1, V^{T}, \Phi(Z)^{T},  (V \otimes \Phi(Z) )^{T} )^{T}, \quad g(X) = (f(X)^{T},  (f(X) \otimes \Phi(Z))^{T} )^{T}.      \end{equation}
We also show some specific forms of $f(X)$ and $g(X)$ for different types of discrete $Z$:
\begin{itemize}
	\item $Z$ is a binary variable,  $f(X) = g(X) =  (1, V^{T}, Z, V^{T} Z )^{T}$.
	\item  $Z$ is trichotomous variable encoded as two dummy variables $(Z_{1}, Z_{2})$,
		$f(X) = g(X) = ( 1, V^{T}, Z_{1},  Z_{2},  V^{T}Z_{1},  V^{T} Z_{2}  )^{T}$.
	\item $Z$ consists of two binary variables $Z_{1}$ and $Z_{2}$, $f(X) = g(X) =	 (1, V^{T}, Z_{1}, Z_{2}, Z_{1}Z_{2},$  $V^{T} Z_{1},  V^{T} Z_{2},  V^{T} Z_{1}Z_{2} )^{T}$.
	\end{itemize}
{It can be seen that the configuration of (\ref{eq21}) will make the dimension of $f(X)$ the same as  $g(X)$}.
  In addition, the proposed setup of $f(X)$ is intuitively sensible, in the sense that  the OR and PS models should include interaction terms between $V$ and $Z$.
		 Proposition \ref{prop2} presents  the large sample properties of    $\hat \mu^{1}(z_{0}; \hat m_{1}, \hat \pi)$ for discrete $Z$, which  can  be deduced from Theorem 2 directly.

\begin{prop}[Doubly robust confidence intervals] \label{prop2}  Suppose that Assumptions  1-2 hold as in Section \ref{sec-theory},  $f(X)$ and $g(X)$ are chosen as in (\ref{eq21}), and
{$( |S_{\bar \gamma}| + |S_{\bar \alpha_1}| ) \log(q)  = o(n^{1/2})$}. Then asymptotic expansion (\ref{eq19}) is valid.
 Moreover, if either PS model (\ref{eq8}) or linear OR model (\ref{eq20}) is correctly specified,   then for any given $z_{0}$, the following results hold for discrete $Z$:

     (i) $n^{1/2} \{ \hat \mu^{1}(z_{0}; \hat m_{1}, \hat \pi)  - {\mu^{1}(z_{0} })  \} \xrightarrow{D}  N \big ( 0, V(z_{0}) \big )$, where $V(z_{0})$ is the same as that in Proposition 1.

   (ii) a consistent estimator of $V(z_{0})$ is $\hat V(z_{0})$, where $\hat V(z_{0})$ is the same as that   in Proposition 1.
		
   (iii) an asymptotic $(1- c)$ confidence interval for {$\mu^{1}(z_{0})$} is $\hat \mu(z_{0}; \hat m_{1}, \hat \pi_{1})
   \pm z_{c/2} \sqrt{ \hat V(z_{0}) / n } $, where $z_{c/2}$ is the $(1- c/2)$ quantile of $N(0, 1)$.
    That is, a doubly robust confidence interval for {$\mu^{1}(z_{0})$} is obtained.
\end{prop}

 It is noteworthy  that asymptotic expansion (\ref{eq19}) holds in Proposition \ref{prop2}  without the need for correctly specified PS model (\ref{eq8}),  while such a result does not hold in Proposition \ref{prop1}.  The reasons for this phenomenon  involve essential ideas about why the proposed methods work. A heuristic interpretation is given in Section \ref{sec-heuristic}.

{For discrete $Z$, stratified analysis is a routinely used method to estimate $\mu^{1}(z)$ (\citep{Abrevaya-Hsu-Lieli2015}). It first splits the sample by $Z$, and then for each subclass, obtains the estimations of $\hat m_{1}$ and $\hat \pi$,  and uses the sample average of $\hat \varphi$ as the estimator of $ \mu^{1}(z)$.}
 Next we  show the connections between the proposed method and stratified analysis for discrete $Z$  and elucidate the advantages of the proposed approach.
 	
  Without of generality, consider the case of binary $Z$,  
 and take $f(X) = g(X) = (1, V^{T}, Z, V^{T} Z )^{T}$ according to (\ref{eq21}).  In order to establish a relationship with stratified analysis,  we rewrite   $f(X)$ as its equivalent expression
     	\begin{equation}     f(X)  = g(X)  = ( I\{Z = 0\}, I\{Z = 0\} V^{T}, I\{Z = 1\}, I\{Z = 1\} V^{T}  )^{T}. \end{equation}
Then by setting the gradient of $L_{CAL}(\gamma)$ and $L_{WL}(\alpha_{1})$ to zero gives that
	\begin{align}
			&   \tilde E \biggl \{  (\frac{T}{ \pi(X; \gamma) } - 1) f(X)     \biggr \} = 0, \label{eq23}   \\	
			&   \tilde E \biggl \{ T \frac {1 - \pi(X; \hat \gamma)} {  \pi(X; \hat \gamma) } (Y - \alpha^{T}_{1} f(X) ) f(X)    \biggr \} = 0.   \label{eq24}
		\end{align}
which are the sample estimating equations for $\gamma$ and $\alpha_{1}$ (up to the lasso penalties in high-dimensional settings). {We focus on analyzing equation (\ref{eq23}), and equation (\ref{eq24}) can be discussed similarly.}  Equation (\ref{eq23}) can be divided into two 
  equations
	\begin{align}
		& \tilde E \biggl \{  (\frac{T}{   [ 1 + \exp\{-\gamma_{0}^{T} f_{0}(X)\}]^{-1}   } - 1)   f_{0}(X)  \biggr \} = 0,
  \label{eq25}		 \\
		& \tilde E \biggl \{  (\frac{T}{  [ 1 + \exp\{-\gamma_{1}^{T} f_{1}(X)\}]^{-1}  } - 1)   f_{1}(X)  \biggr \} = 0,  \label{eq26}
	\end{align}
 where $f_{0}(X) = I\{Z = 0\} (1, V^{T})^{T}$, $f_{1}(X) = I\{Z = 1\} (1, V^{T})^{T}$,  $\gamma = (\gamma_{0}^{T}, \gamma_{1}^{T})^{T}$ that satisfies $\gamma^{T} f(X) = \gamma_{0}^{T} f_{0}(X) +  \gamma_{1}^{T} f_{1}(X)$.  Equations (\ref{eq25}) and (\ref{eq26}) are equivalent to
 the sample estimating equations in stratified analysis.
 	 However,  if there are multiple categories,  then stratified analysis is {troublesome}, especially in high-dimensional settings,
 where stratified analysis may select different tuning parameters for lasso penalties and different covariates in different strata.
 The proposed method is numerically more tractable with only two lasso tuning parameters for the PS and OR models.

\subsection{Estimations of $\mu^{0}(z)$ and $\tau(z)$.}    \label{sec-3.4}
The  results presented in Propositions \ref{prop1} and \ref{prop2}  mainly focus on estimation of $\mu^1(z)$, but they can be directly extended for estimating  $\mu^{0}(z)$ and $\tau(z)$.
{Similar to (\ref{eq16}),  we posit a marginal structural model for $\mu^{0}(z)$ based on basis functions $(1, \Phi(z))$.}

  In addition to the propensity score model (\ref{eq8}) and generalized linear outcome model (\ref{eq2}), consider the following outcome regression model in the untreated population,
		\begin{equation}  \label{eq27}
		   E(Y | T = 0, X) = m_{0}(X; \alpha_{0}) = \psi\{  \alpha_{0}^{T} g(X)  \},       \end{equation}
where $g(X)$ is the same 
as in model (\ref{eq2}) and $\alpha_{0}$ is a vector of unknown parameters.  Then for a given $z_{0}$, our point estimator of {$\tau(z_{0})$} is $\hat \mu^{1}(z_{0}; \hat m_{1}, \hat \pi) - \hat \mu^{0}(z_{0}; \hat m_{0}, \hat \pi_{0})$, with
	\[    \hat \mu^{0}(z_{0}; \hat m_{0}, \hat \pi_{0}) =    \Phi^\dag(z_{0})^{T}  \tilde E^{-1} \big [    \Phi^\dag(Z)    \Phi^\dag(Z)^{T}    \big ]     \tilde E \big [   \Phi^\dag(Z)   \varphi(Y, 1- T, X; \hat m_{0}, 1 - \hat \pi_{0} )          \big ],        \]
where $\varphi(\cdot)$ is defined in (\ref{eq6}),  $\hat \pi_{0} = \pi(X; \hat \gamma_{0, RCAL})$,  $\hat m_{0} = m_{0}(X; \hat \alpha_{0, RWL})$, and
 $\hat \gamma_{0, RCAL}$ is defined as a minimizer of (\ref{eq9}), but with the calibration loss function $L_{CAL}(\gamma)$ replaced by
		$        L_{0, CAL}(\gamma) = \tilde E\{  (1- T) e^{\gamma^{T}  f(X)} -  T \gamma^{T} f(X)    \}.                  $
The estimator $\hat \alpha_{0, RWL}$ is defined as a minimizer of
		$   L_{0, RWL}(\alpha_{0}; \hat \gamma_{0, RCAL}) =  L_{0, WL}(\alpha_{0}; \hat \gamma_{0, RCAL})    +     \lambda || (\alpha_{0})_{1:q} ||_{1},               $
with
		\[ L_{0, WL}(\alpha_{0}; \hat \gamma_{0, RCAL})   = \tilde E\big [ \frac{ 1 - T  }{ w(X; \hat \gamma_{0, RCAL} )} \{ - Y\alpha_{0}^{T} g(X) + \Psi( \alpha_{0}^{T} g(X) )    \}      \big ].    \]
Under similar conditions in Proposition \ref{prop1} or \ref{prop2},  the estimator $\hat \mu^{0}(z_{0}; \hat m_{0}, \hat \pi_{0})$ admits an asymptotic expansion
	\begin{equation}
			  \hat \mu^{0}(z_{0}; \hat m_{0}, \hat \pi_{0}) =  \hat \mu^{0}(z_{0}; \bar m_{0}, \bar \pi_{0})  + o_{p}(n^{-1/2}),
	\end{equation}
where $\bar \pi_{0} = \pi(X; \bar \gamma_{0})$, $\bar m_{0} = m_{0}(X; \bar \alpha_{0})$,  $\bar \gamma_{0}$ and $\bar \alpha_{0}$ are defined as the minimizers of $E\{ L_{0, CAL}(\gamma)  \}$ and  $E\{ L_{0, WL}(\alpha_{0}; \hat \gamma_{0, RCAL}) \}$, respectively.  In particular, an asymptotic $(1- c)$ confidence interval for {$\tau(z_{0})$} can be given as
		\begin{equation} \label{eq29}  \hat \mu^{1}(z_{0}; \hat m_{1}, \hat \pi) - \hat \mu^{0}(z_{0}; \hat m_{0}, \hat \pi_{0})       \pm z_{c/2} \sqrt{ \mathbb{\hat  V}(z_{0}) / n },                 \end{equation}
where
	$ \mathbb{\hat V}(z_{0})  =      \Phi^\dag(z_{0})^{T}  M^{-1}  \mathbb{ \hat  G}   M^{-1}     \Phi^\dag(z_{0})  / n,   $
		$\mathbb{\hat G}  = \tilde E  \big \{    \Phi^\dag(Z)    \Phi^\dag(Z)   [
		\hat \varphi_{\tau}   -	(\tilde \beta_{0}, \tilde \beta_{1}^{T} )   \Phi^\dag(Z)    ]^{2}      \big \},
		 $
 $M =  \tilde E \{    \Phi^\dag(Z)    \Phi^\dag(Z)^{T}  \}$,
 $\hat \varphi_{\tau} = \varphi( Y, T, X; \hat m_{1}, \hat \pi)  - \varphi( Y, 1 - T, X; \hat m_{0}, 1 - \hat \pi_{0})$,  and
$    (\tilde \beta_{0}, \tilde \beta_{1}^{T})^{T}  =  \tilde E^{-1} \{   \Phi^\dag(Z)   \Phi^\dag(Z)^{T}     \} \tilde E[   \Phi^\dag(Z)  \hat \varphi_{\tau}  ].$



\section{Asymptotic properties}  \label{sec-asymptotic}

\subsection{Heuristic discussion}  \label{sec-heuristic}
We delineate basic ideas underlying the construction of the  estimators  $\hat \gamma$ and $\hat \alpha_{1}$, and point out  why we need careful specification of $f(X)$ and $g(X)$ in (\ref{eq18}) or (\ref{eq21}), such that the estimator $\hat \mu^{1}(z_{0}; \hat \pi_{1}, \hat \pi)$ satisfies asymptotic expansion (\ref{eq19}), under possible model misspecification. The discussion here is  heuristic, and formal theory is presented in Section \ref{sec-theory}.
For a given $z_{0}$,
	\[     \hat \mu^{1}(z_{0};  \hat m_1, \hat \pi )  =  \hat \mu^{1}(z_{0};  \bar m_{1}, \bar \pi )    + \Phi^\dag(z_{0})^{T}  \tilde E^{-1} \{ \Phi^\dag(Z) \Phi^\dag(Z)^{T}  \}  \tilde E\{  \Phi^\dag(Z)  (\hat \varphi -  \bar \varphi )     \}.          \]
Under mild assumptions, for (\ref{eq19}) to hold, it suffices to show that
		\begin{equation}  \label{eq30}
		\tilde E\{  \Phi^\dag(Z)  (\hat \varphi -  \bar \varphi )    = o_{p}(n^{-1/2}).
		\end{equation}				
By a Taylor expansion of $\tilde E\{  \Phi^\dag(Z) \hat \varphi  \}$,
 	\begin{equation}  \label{31}
			\tilde E\{ \Phi^\dag(Z) \hat \varphi  \}	= \tilde E\{  \Phi^\dag(Z)  \bar \varphi \} + (\hat \alpha_{1} - \bar \alpha_{1})^{T}   \Delta_{1}  + (\hat \gamma - \bar \gamma)^{T} \Delta_{2} +  o_{p}(n^{-1/2}),
	\end{equation}
 where the remainder is taken to be $o_{p}(n^{-1/2})$ under suitable conditions, and 	
 	\begin{align*}
	 \Delta_{1}  ={}&     \frac{  \partial }{\partial \alpha_{1} } \tilde E(   \Phi^\dag(Z)  \varphi(Y, T, X;  \alpha_{1}, \gamma)  ) \Big |_{(\alpha_{1}, \gamma) = (\bar \alpha_{1}, \bar \gamma)}, \\
	 \Delta_{2}  ={}&    \frac{  \partial }{\partial \gamma } \tilde E(    \Phi^\dag(Z)  \varphi(Y, T, X;  \alpha_{1}, \gamma)  ) \Big |_{(\alpha_{1}, \gamma) = (\bar \alpha_{1}, \bar \gamma)}.
	\end{align*}	
 To show (\ref{eq30}), it is sufficient to show that $ (\hat \alpha_{1} - \bar \alpha_{1})^{T} \Delta_{1} = o_{p}(n^{-1/2})$ and $  (\hat \gamma - \bar \gamma)^{T} \Delta_{2} = o_{p}(n^{-1/2})$  with possible model specification.    In general,  $\hat \alpha_{1} - \bar \alpha_{1}$ and $\hat \gamma - \bar \gamma$ are no smaller than $O_{p}(n^{-1/2})$ in low- or high-dimensional settings. In order to get the desired convergence rates,  the crucial point is that $\Delta_{1}$ and $\Delta_{2}$ should be $o_{p}(1)$,  and their corresponding population version should satisfy
 			 	\begin{align}
	  \frac{  \partial }{\partial \alpha_{1} }  E(   \Phi^\dag(Z) \varphi(Y, T, X;  \alpha_{1}, \gamma)  ) \Big |_{(\alpha_{1}, \gamma) = (\bar \alpha_{1}, \bar \gamma)} ={}& 0 , \label{eq32} \\
   \frac{  \partial }{\partial \gamma }  E(    \Phi^\dag(Z)  \varphi(Y, T, X;  \alpha_{1}, \gamma)   ) \Big |_{(\alpha_{1}, \gamma) = (\bar \alpha_{1}, \bar \gamma)} ={}& 0 . \label{eq33}
	\end{align}	
Hence  a natural approach is to solve (\ref{eq32}) and (\ref{eq33}) being in low-dimensional settings and add lasso penalties in high-dimensional settings. Nevertheless, this method will encounter
with a basic problem: there are more equations than parameters.
	{It is easy to see that (\ref{eq32}) includes $(K+1)(q+1)$ equations and (\ref{eq33}) contains $(K+1)(p+1)$ equations, while the dimensions of $\gamma$ and $\alpha_{1}$ are $p+1$ and $q+1$, respectively. Therefore, the coefficients  $\gamma$ and $\alpha_{1}$  cannot be identified by solving  (\ref{eq32}) and (\ref{eq33}) without further consideration.}
Fortunately, this difficulty can be overcome by simply a careful specification of $f(X)$ and $g(X)$.

Specifically, with 	PS model (\ref{eq8}) and linear OR model (\ref{eq20}),  $\Delta_{1}$ and $\Delta_{2}$ reduces to
 	\begin{align*}
	   \Delta_{1}  ={}&  \tilde E \biggl \{  (\frac{T}{ \bar \pi(X) } - 1) g(X)  \otimes \Phi^\dag(Z)     \biggr \} ,     \\	
	 \Delta_{2}  ={}&    \tilde E \biggl \{ T \frac{1 -  \bar \pi(X)}{  \bar \pi(X) } (Y - \bar \alpha_{1}^{T} g(X) ) f(X) \otimes \Phi^\dag(Z)    \biggr \}.
	\end{align*}
If $g(X)$ satisfies the form of (\ref{eq18}), then according to the definition of $\bar \alpha_{1}$,  (\ref{eq33}) holds regardless of whether the OR model is specified correctly.
 In addition,  (\ref{eq32}) holds provided that PS model (\ref{eq8}) is correctly specified but OR model (\ref{eq20}) may be misspecified, {which  elucidates why Proposition \ref{prop1} can be derived.}
Furthermore,  if $f(X)$ and $g(X)$ are specified as in (\ref{eq21}),  then  $\Delta_{1}$ and $\Delta_{2}$ have a  simpler form with discrete $Z$:
		\begin{align*}
	   \Delta_{1}  ={}&  \tilde E \biggl \{  (\frac{T}{ \bar \pi(X) } - 1) f(X)     \biggr \} ,     \\	
	 \Delta_{2}  ={}&    \tilde E \biggl \{ T \frac{1 -  \bar \pi(X)}{  \bar \pi(X) } (Y - \bar \alpha_{1}^{T} g(X) ) g(X)    \biggr \},
	\end{align*}	
which exactly are the gradients of  $L_{CAL}(\bar \gamma)$ and $L_{WL}(\alpha_{1}; \bar \gamma)$, respectively. In this case,  (\ref{eq32}) and (\ref{eq33}) hold  just by the definition of $\bar \gamma$ and $\bar \alpha_{1}$, irrespective of the model specifications for PS and OR, {which explains why Proposition \ref{prop2} can be obtained.}


\subsection{Theoretical analysis}  \label{sec-theory}

First, we summarize the results which can be deduced directly from \citet{Tan-Annals2020} about  $(\hat \gamma, \hat \alpha_{1})$.
   For a matrix $\Sigma$ with row indices $\{0, 1, ..., k \}$, a compatibility condition condition (\citep{Buhlmann-VanDeGeer2011}) is said to hold with a subset $S \in \{ 0, 1, ..., k\}$ and constants $\nu > 0$ and $\xi > 1$ if $\nu^{2} (  \sum_{j \in S} | b_{j} |^{2} \leq b^{T} \Sigma b ) $ for any vector $b = (b_{0}, b_{1}, ..., b_{k}) \in \mathbb{R}^{k+1}$ satisfying $ \sum_{j\notin S} | b_{j} | \leq \xi \sum_{j \in S} | b_{j}|$.

\bigskip
\noindent
{\bf Assumption 1.} Suppose that the following conditions are satisfied:
	\begin{itemize}
	\item[(i)]  $max_{j=0, 1, ..., p} | f_{j}(X) | \leq C_{0}$ almost surely for a constant $C_{0} \geq 1$;
		\item[(ii)] $ \bar \gamma^{T}  f(X) \geq B_{0}$ almost surely for a constant $B_{0}$, that is, $\pi(X; \bar \gamma) \geq (1+ e^{-B_{0}})^{-1}$.
	\item[(iii)] a compatibility  condition holds for $\Sigma_{f}$ with the subset $S_{\bar \gamma} = \{ 0\} \cup \{j: \bar \gamma_{j} \neq 0, j = 1, ..., p  \}$ and some constants $\nu_{0} >0$ and $\xi_{0} > 1$, where  $\Sigma_{f} = E[  T w(X; \bar \gamma) f(X) f(X)^{T}  ]$ is the Hessian of $E\{ L_{CAL}(\gamma) \}$  at $\gamma = \bar \gamma$ and $w(X; \bar \gamma)  = e^{-\gamma^{T} f(X)}$.
	 \item[(iv)]   $| S_{\bar \gamma} | \lambda_{0} $ is sufficiently small.
\end{itemize}

\bigskip
\noindent
{\bf Assumption 2.} Suppose that the following conditions are satisfied:
	\begin{itemize}	
		\item[(i)]  $max_{j=0, 1, ...,  q} | g_{j}(X) | \leq C_{1}$ almost surely for a constant $C_{1} \geq 1$;
			\item[(ii)] $ \bar \alpha_{1}^{T} g(X)$ is bounded in absolute values by $B_{1} > 0$ almost surely;    		 \item[(iii)]  $\psi'(u) \leq \psi'(\tilde u) e^{C_{2} | u - \tilde u |}$ for any $(u, \tilde u)$, where $C_{2}$ is a constant.
	 \item[(iv)]   $Y^{1} - m_{1}(X; \bar \alpha_{1})$ is uniformly sub-Gaussian given $X$:
	 			$$ D_{0}^{2} E \big [ \exp\{  (Y^{1} -  m_{1}(X; \bar \alpha_{1}) )^{2} / D_{0}^{2}  \}  - 1  \big | X \big ] \leq D_{1}^{2} $$ for some positive constants $(D_{0}, D_{1})$.
		\item[(v)]  a compatibility condition holds for $\Sigma_{g}$ with the subset $S_{\bar \alpha_{1}} = \{ 0 \} \cup \{j: \bar \alpha_{1, j} \neq 0, j = 1, ..., p  \}$ and some constants $\nu_{1} >0$ and $\xi_{1} > 0$, where $\Sigma_{g} = E[T w(X; \bar \gamma) g(X) g(X)^{T} ]$.
		\item[(vi)]  $ | S_{\bar \gamma} | \lambda_{0} + | S_{\bar \alpha_{1}} | \lambda_{1}  $ is sufficiently small.
	\end{itemize}

Theorem \ref{th1} summarizes the results of \citet{Tan-Annals2020} related to $(\hat \gamma, \hat \alpha_{1})$.

\begin{thm}[\citep{Tan-Annals2020}] \label{th1}  Suppose Assumptions 1 and 2 hold and $\lambda_{0} \leq 1$. Then for sufficiently large constants $A_{0}$ and $A_{1}$.

(a)  If OR model (\ref{eq2}) is used, {$g(X)$ is specified as in (\ref{eq18})},  then we have with probability at least $1 - c_{0} \epsilon$,
		\begin{align}
	  & 	|| \hat \gamma - \bar \gamma ||_{1}  \leq  M_{0} | S_{\bar \gamma} | \lambda_{0},  \quad ( \hat \gamma - \bar \gamma )^{T} \tilde \Sigma_{f}  ( \hat \gamma - \bar \gamma )  \leq M_{0} | S_{\bar \gamma} | \lambda_{0}^{2}      \label{eq34}  \\
 	& 	    || \hat \alpha_{1} - \bar \alpha_{1} ||_{1}  \leq  M_{1}  (  | S_{\bar \gamma} | \lambda_{0} + | S_{\bar \alpha_{1}} | \lambda_{1}) ,    \quad ( \hat \alpha_{1} - \bar \alpha_{1} )^{T}  \tilde \Sigma_{g}  ( \hat \alpha_{1} - \bar \alpha_{1} )  \leq  M_{1}  (  | S_{\bar \gamma} | \lambda_{0}^{2} + | S_{\bar \alpha_{1}} | \lambda_{1}^{2} ),    \label{eq35}
		\end{align}
where  $c_{0}$, $M_{0}$ and $M_{1}$ are positive constants, $\tilde \Sigma_{f}$ and $\tilde \Sigma_{g}$ are the sample versions of $\Sigma_{f}$ and $\Sigma_{g}$, i.e., $\tilde \Sigma_{f} = \tilde E[  T w(X; \bar \gamma) f(X) f(X)^{T}  ]$ and $\tilde \Sigma_{g} = \tilde E[  T w(X; \bar \gamma) g(X) g(X)^{T}  ]$.  Furthermore, if PS  model (\ref{eq8}) is correctly specified,
we also have with probability at least  $1 - c_{0} \epsilon$,
	\begin{equation}  \label{eq36}
			\big |  \tilde E( \hat \varphi  - \bar \varphi )   \big | \leq M_{2} (  | S_{\bar \gamma} | \lambda_{0}  +  | S_{\bar \alpha_{1}} | \lambda_{1} ) \lambda_{1},	
	\end{equation}
where $M_{2}$ is a positive constant.

 (b) If {linear OR model (\ref{eq20}) is used}, {$f(X)$ and $g(X)$ are specified as in (\ref{eq21})}, then the results (\ref{eq34}), (\ref{eq35}) and (\ref{eq36}) also hold.
\end{thm}

Inequalities (\ref{eq34}) and (\ref{eq35}) lead directly to the convergence rates for $(\hat \gamma, \hat \alpha_{1})$,
	\begin{align*}
		&	|| \hat \gamma - \bar \gamma ||_{1} = O_{p}(1) \cdot | S_{\bar \gamma} | \{  \log(p) / n   \}^{1/2}, \\
		&	|| \hat \alpha_{1} - \bar \alpha_{1} ||_{1} = O_{p}(1) \cdot ( | S_{\bar \gamma} | + | S_{\bar \alpha_{1}} | ) \{  \log(q) / n   \}^{1/2}.
	\end{align*}
Inequality (\ref{eq36}) will be used to establish the inequalities (\ref{eq37}) in Theorem 2(a) and (\ref{eq39}) in Theorem 3(a),
two crucial results in this article.
The following Theorem \ref{th-discrete} presents  the  large sample properties of
the proposed estimator $\hat \mu^{1}(z; \hat m_{1}, \hat \pi)$ for discrete $Z$.

\begin{thm}[doubly robust confidence intervals] \label{th-discrete}
Under the conditions of Theorem 1(b),  then for any given $z_{0}$ of discrete $Z$, we have

  (a)  with probability at least $1 - c_{0} \epsilon$,
		\begin{equation} \label{eq37}
	 \big  |    \hat \mu^{1}(z_{0}; \hat m_{1}, \hat \pi)  -  \hat \mu^{1}(z_{0}; \bar m_{1}, \bar \pi)     \big |  \leq M_{3} (  | S_{\bar \gamma} | \lambda_{0}  +  | S_{\bar \alpha_{1}} | \lambda_{1} ) \lambda_{1},	
		\end{equation}
where $M_{3}$ is a positive constant.

 (b) with probability {$1 - (c_{0} +4 ) \epsilon$},
		\begin{equation} \label{eq38}
			 \hat V(z_{0})  -  V(z_{0})  = o_{p}(1),
		\end{equation}
provided that  {$ ( |S_{\bar \gamma} + |S_{\bar \alpha_{1}}|  ) \sqrt{\log(q)}  = o(n^{1/2})$}, where $V(z_{0})$ and $\hat V(z_{0})$ are defined in Proposition 2(i)--(ii). 	

\end{thm}

{It should be noted that $\hat \mu^{1}(z; \bar m_{1}, \bar \pi)$ is a doubly robust point estimator of $\mu^{1}(z)$. Therefore, (\ref{eq37}) implies that  $\hat \mu^{1}(z; \hat m_{1}, \hat \pi)$ is also a doubly robust point estimator of $\mu^{1}(z)$, provided that $(  | S_{\bar \gamma} | \lambda_{0}  +  | S_{\bar \alpha_{1}} | \lambda_{1} ) \lambda_{1} = o(1)$, that is, $(  |S_{\bar \gamma}| + |S_{\bar \alpha_1}| ) \log(q)  = o(n)$.  In addition, to obtain a valid confidence intervals, it requires the asymptotic expansion (\ref{eq19}) to hold, which implies that  $( | S_{\bar \gamma} | \lambda_{0}  +  | S_{\bar \alpha_{1}} | \lambda_{1} ) \lambda_{1} = o(n^{-1/2})$, namely, $(  |S_{\bar \gamma}| + |S_{\bar \alpha_1}| ) \log(q)  = o(n^{1/2})$. In summary,}
Theorem \ref{th-discrete} shows that, for discrete $Z$ with linear OR model (\ref{eq20}) and specification of $f(X)$ and $g(X)$  as in (\ref{eq21}), the proposed method obtains both doubly point estimators and doubly confidence intervals for $\mu^{1}(z)$, provided that {$( |S_{\bar \gamma}| + |S_{\bar \alpha_1}| ) \log(q)  = o(n^{1/2})$}, which leads to Proposition \ref{prop2}.  Similar to Theorem  \ref{th-discrete}, the following Theorem \ref{th-continuous}  implies the results presented  in Proposition \ref{prop1}.

\begin{thm}[Model-assisted confidence intervals] \label{th-continuous}  Under the conditions of {Theorem 1(a)}, if PS model (\ref{eq8}) is correctly specified,  then for a given value $z_{0}$ of discrete $Z$ or continuous $Z$, we have

	(a) with probability at least {$1 - (c_{0}+8) \epsilon$},
				\begin{equation}  \label{eq39}
	 \big  |    \hat \mu^{1}(z_{0}; \hat m_{1}, \hat \pi)  -  \hat \mu^{1}(z_{0}; \bar m_{1}, \bar \pi)     \big |  \leq M_{4} (  | S_{\bar \gamma} | \lambda_{0}  +  | S_{\bar \alpha_{1}} | \lambda_{1} ) \lambda_{1},	
		\end{equation}
	where $M_{4}$ is a  positive constant. 		 	

  (b) with probability at least {$1 - (c_{0}+12) \epsilon$},
		\begin{equation}   \label{eq40}
	 \hat V(z_{0})  -  V(z_{0})  = o_{p}(1),
		\end{equation}
	provided that  {$ ( |S_{\bar \gamma} + |S_{\bar \alpha_{1}}|  ) \sqrt{\log(q)}  = o(n^{1/2})$}, 	where $V(z_{0})$ and $\hat V(z_{0})$ are defined in Proposition 1(i)--(ii).
	
\end{thm}

The preceding theoretical analysis focuses on the large sample properties of $\hat \mu^{1}(z; \hat m_{1}, \hat \pi)$. Similar results can be derived 
  for  $\hat \mu^{0}(z; \hat  m_{0}, \hat \pi_{0})$ and $\hat \mu^{1}(z; \hat m_{1}, \hat \pi) - \hat \mu^{0}(z; \hat  m_{0}, \hat \pi_{0})$ by analogous arguments.


\section{Simulation studies} \label{sec-simu}
Extensive simulation studies are carried out to evaluate the finite sample performance of the
proposed methods. We consider three scenarios of $Z$: binary variable $Z$, continuous variable $Z$,  and $Z$ consists of two binary variables $Z_{1}$ and $Z_{2}$.  The regularized calibrated estimation for PS model and regularized weighted likelihood estimation for OR model can be implemented by using R package {\bf RCAL} (\citep{Tan2019}), and the corresponding tuning parameters are determined via using 5-fold cross validation. 

Throughout this simulation,  the data generating processes of covariates are as follows:
    $V = (V_{1}, ..., V_{d}) \sim Normal(0, \Sigma)$ with $\Sigma_{j, k} = 2^{-|j - k|}$ for $1 \leq j, k \leq d$, and {independently}, $Z \sim Binomial(1, 0.5)$ or $Z \sim Uniform(-0.5, 0.5)$ for discrete or continuous $Z$.
    For $Z$ consisting of two binary variables $(Z_{1}, Z_{2})$,  $Z_{1}$ and $Z_{2}$ are independent and identically distributed from  $Binomial(1, 0.5)$. The error term is $\epsilon \sim Normal(0, 1)$.
     Let $\gamma =0.5  (1, -1, -1, 1, -1)^{T}$,  $X = (Z^{T}, V^{T})^{T}$ and $V_{i}$ be $i$-th element of $V$.

\bigskip
  {\bf Discrete $Z$}.  We first consider the following three different data generating scenarios (C1)-(C3) with binary $Z$. The three cases can help to assess the doubly robust properties
  for both point estimators and confidence intervals. 

	\begin{enumerate}
	   \item[(C1)] $Z$ is binary variable,
	   $P(T=1|X) =  \{1 +  \exp(  -(Z, V_{1}, V_{2}, V_{3}, V_{4})^{T} \gamma ) \}^{-1}$,
	   $Y^{1} = 1 + Z + \sum_{i=1}^{4} \{ V_{i} Z + 2 V_{i} (1 - Z) \} + \epsilon$.
	
  \item[(C2)]
 Generate	$Z, V$ and $T$ as in case (C1),
  $Y^{1} = 1 + Z + \sum_{i=1}^{4} \{ V_{i} Z + 2 V_{i} (1 - Z) + V_{i}^{3}/ 2^{i} \}  +   \epsilon$.

	   \item[(C3)]
	   Generate	$Z, V$ and $Y^{1}$ as in case (C1),
	  $P(T=1|X) = \{1 +  \exp( - (Z, V_{1}^{2}, V_{2}^{2}, V_{3}^{2}, V_{4}^{2} )^{T} \gamma ) \}^{-1}$.
	
	\end{enumerate}
The three scenarios  can be classified as follows:
	\begin{enumerate}
	   \item[(C1)]    PS and OR models both correctly specified.
	   \item[(C2)]    PS model correctly specified, but OR model misspecified.
	   \item[(C3)]    PS model misspecified, but OR model correctly specified.
	\end{enumerate}
The true curve of $\mu^{1}(z)$ is $1+ z$ for  all cases of (C1)-(C3).
   We set $f(X) = g(X) = (1, V^{T}, Z, V^{T} Z)^{T}$  as discussed in Section \ref{doubly-robust}.
Each simulation study is based on 1000 replicates  with sample size $n=500$.
 Bias and Var are the Monte Carlo bias and variance  over the  1000 simulations of the points estimates. EVar is the mean of the variance estimates. Cov90 and Cov95 are the coverage proportions of the 90\% and 95\%   confidence intervals  by using the asymptotic variance formula, respectively.
  Table \ref{tab1} summarizes the results of  $\hat \mu^{1}(z)$  for scenarios  (C1)-(C3).

%
%
%


\begin{table}[H]
\caption{Estimations of $\mu^{1}(z)$  for binary variable $Z$}
\centering
\footnotesize
 \label{tab1}
\begin{tabular}{c  rcccc  c rcccc}
		\hline
		  &    \multicolumn{5}{c}{  $n = 500$, $p = 200$   }  & &     \multicolumn{5}{c}{  $n = 500$, $p = 400$  }                               \\
		    $\hat \mu^{1}(z)$   &  Bias & $\sqrt{\text{Var}}$   &   $\sqrt{\text{EVar}}$ & Cov90 & Cov95 & &  Bias & $\sqrt{\text{Var}}$   &  $\sqrt{\text{EVar}}$ & Cov90 & Cov95  \\
		\hline		
		  &  \multicolumn{11}{c}{ 	\multirow{2}{*}{(C1) cor PS, cor OR  } }  \\
		  &  \multicolumn{11}{c}{} \\
	   $\hat \mu^{1}(0)$	  &  -0.032 & 0.370 & 0.371 & 0.905 & 0.954  &  & -0.034 & 0.366 & 0.368 & 0.897 & 0.950   \\
			  $\hat \mu^{1}(1)$ & -0.040 & 0.201 & 0.200 & 0.879 & 0.952 & & -0.038 & 0.202 & 0.200 & 0.889 & 0.943   \\
			  	  &  \multicolumn{11}{c}{  \multirow{2}{*}{(C2) cor PS, mis OR  } }  \\
				 &  \multicolumn{11}{c}{} \\
	   $\hat \mu^{1}(0)$	 & -0.054 & 0.516 & 0.501 & 0.896 & 0.944	& & -0.063 & 0.515 & 0.498 & 0.884 & 0.941  \\
			  $\hat \mu^{1}(1)$ & -0.081 & 0.348 & 0.337 & 0.887 & 0.935  & & -0.072 & 0.336 & 0.336 & 0.898 & 0.951     \\
			  			  	  &  \multicolumn{11}{c}{  \multirow{2}{*}{(C3) mis PS, cor OR  } }  \\
				 &  \multicolumn{11}{c}{} \\
	   $\hat \mu^{1}(0)$	 & 0.019 & 0.377 & 0.361 & 0.888 & 0.938  & & 0.033 & 0.375 & 0.357 & 0.874 & 0.935   \\
			  $\hat \mu^{1}(1)$ & -0.003 & 0.195 & 0.202 & 0.905 & 0.955  & & -0.016 & 0.203 & 0.200 & 0.888 & 0.947 	     \\
		\hline			
	\end{tabular}
	\begin{flushleft}
	Note: Both the dimensions of $f(X)$ and $g(X)$ are $p+1$.
	\end{flushleft}
\end{table}
\normalsize
%
%

As shown in Table \ref{tab1}, for all the cases (C1)-(C3), the Bias is small,  $\sqrt{\text{EVar}}$  is close to $\sqrt{\text{Var}}$ and  the coverage proportions Cov90 and Cov95 are around the nominal levels 0.90 and 0.95, respectively.  Because case (C2) involves a misspecified OR model and case (C3) involves  a misspecified PS model,  the results of cases (C2) and (C3)  justify that  both the point estimators and confidence intervals
     are doubly robust.
%

\bigskip
{\bf Continuous $Z$}.
We consider two data generation mechanisms {with} continuous $Z$: 
	\begin{enumerate}
			\item[(C4)]    $Z$ is a continuous variable,  $P(T=1|X) = \{1 +  \exp( - (Z, V_{1}, V_{2}, V_{3}, V_{4})^{T} \gamma  ) \}^{-1}$,     $Y^{1} = Z + \sum_{i=1}^{4}  V_{i}  + \epsilon$.
	   				
			\item[(C5)]    Generate	$Z, V$ and $T$ as in case (C4),
  $Y^{1} =  Z(1+2Z)^2(Z-1)^2  + \sum_{i=1}^{4} ( V_{i}^{2} + V_{i} ) / 2^{i+1}   + \epsilon$.   		
    	\end{enumerate}

 Scenario (C4) implies $\mu^{1}(z)$ is a linear function of $z$, while scenario (C5) indicates
 $\mu^{1}(z)$ is a {polynomial function}  of $z$.
  The vectors $f(X)$ and $g(X)$ are specified as follows:
 	$f(X) = (1, V^{T}, Z)^{T}, ~ g(X) = ( 1, V^{T}, Z, \Phi(Z)^{T}, V^{T} \phi_{1}(Z), ..., V^{T} \phi_{K}(Z),  \Phi(Z)^{T} Z )^{T}.$  In this simulation,  we let $\Phi(Z)$ be cubic spline basis functions with three knots selected by the 25\%, 50\% and 75\% sample quantiles of $Z$, which is a six-dimensional  random vector excluding a constant and can be implemented using R package {\bf gam} (\citep{Hastie2018}).
Since $Z$ is continuous for cases (C4) and (C5), we report the simulation results at five representative points of $Z:  -0.4, -0.2, 0, 0.2, 0.4$. 
Table \ref{tab2} shows the numerical results of $\hat \mu^{1}(z)$ for cases (C4) and (C5). 
 Both of them have a similar performance with those of discrete $Z$.

\begin{table}[H]
\caption{Estimations of $\mu^{1}(z)$ for continuous $Z$, $n=500$, $p=60$, $q = 420$}
\centering
\footnotesize
 \label{tab2}
\begin{tabular}{c  rcccc  c rcccc}
		\hline
		    $\hat \mu^{1}(z)$   &  Bias & $\sqrt{\text{Var}}$   &   $\sqrt{\text{EVar}}$ & Cov90 & Cov95 & &  Bias & $\sqrt{\text{Var}}$   &  $\sqrt{\text{EVar}}$ & Cov90 & Cov95  \\
		\hline		
		  &  \multicolumn{5}{c}{  \multirow{1}{*}{(C4) cor PS, cor OR} }  & &   \multicolumn{5}{c}{  \multirow{1}{*}{(C5) cor PS, mis OR} }    \\
	    $\hat \mu^{1}(-0.4)$   & 0.011 & 0.409 & 0.400 & 0.886 & 0.942     & & -0.027 & 0.201 & 0.199 & 0.882 & 0.942 \\
			 $\hat \mu^{1}(-0.2)$    & -0.035 & 0.359 & 0.341 & 0.881 & 0.936 & &  -0.020 & 0.172 & 0.170 & 0.884 & 0.944 \\
	    $\hat \mu^{1}(0.0)$  	& -0.036 & 0.339 & 0.331 & 0.892 & 0.940   & & -0.025 & 0.179 & 0.166 & 0.866 & 0.928\\
			 $\hat \mu^{1}(0.2)$   & -0.013 & 0.341 & 0.342 & 0.899 & 0.948  &   &  -0.036 & 0.173 & 0.177 & 0.894 & 0.948 \\
			  $\hat \mu^{1}(0.4)$   & -0.034 & 0.414 & 0.403 & 0.883 & 0.941  &  & -0.028 & 0.208 & 0.207 & 0.880 & 0.938 \\
	\hline			
	\end{tabular}
	\begin{flushleft}
	Note:  the {dimensions} of $f(X)$ and $g(X)$ are $p+1$ and $q+1$ , respectively.
	\end{flushleft}
\end{table}
\normalsize




The setups for all preceding numeric results are in exact sparsity settings. A more common scenario in modern applications is approximate sparsity, which means that all covariates are relevant associated with nonzero coefficients but only a few are truly important with large coefficients.
We also conduct  numeric experiments to assess the finite sample performance of proposed methods under  approximate sparsity settings {for both binary $Z$, continuous $Z$ and $Z$ consisting of two binary variables}. The corresponding results are similar to those of in exact sparsity settings and are presented in Table {\color{blue}S1} of the supplementary material.

Finally, we compare the proposed method  with competing AIPW methods of  \citet{Fan-Hsu-Lieli-Zhang2019} and \citet{Zimmert-Lechner2019} discussed in Section \ref{sec-existing-estimators}  for continuous $Z$. We adopt the AIPW methods with full sample and four-fold cross-fitting as suggested in  \citet{Fan-Hsu-Lieli-Zhang2019}.  For PS and OR models, we set  $f(X) = g(X) = (1, V^{T}, Z)^{T}$ and the associated  tuning parameters are selected by 5-fold cross validation.  As done in \citet{Fan-Hsu-Lieli-Zhang2019},   we use the Gaussian kernel throughout and the bandwidth is set as  $h = \hat h_{opt} \times n^{1/5} \times n^{-2/7}$, where $\hat h_{opt}$  is  a plug-in estimator of  the optimal bandwidth that can be implemented using R package {\bf KernSmooth} (\citep{Ruppert-etal1995}; \citep{Wand2015}).

Table \ref{tab3} summarizes the results of competing AIPW estimators for cases (C4) and (C5).
 When both PS and OR models are correctly specified,  the two competing AIPW methods  perform well in terms of Bias, Cov90 and Cov95.  Nevertheless, when OR model is misspecified,  their
  coverage proportions are away from the nominal values. This indicates that the competing approaches  do not enjoy the property of doubly robust confidence intervals. 
   As expected, the AIPW estimator with full sample tends to have larger Bias and smaller  $\sqrt{\text{Var}}$ than that of four-fold cross-fitting, in that sample-splitting may decrease bias and induce random errors in finite sample.
   In addition, by comparison of Table \ref{tab3} with Table \ref{tab2}, the proposed method has similar performance with competing AIPW methods 
     when both PS and OR models are correctly specified. However, with a misspecified OR model,  the proposed approach  has smaller Bias and $\sqrt{\text{Var}}$, and better coverage proportions.

\begin{table}[H]
\caption{Comparison of competing approaches for continuous $Z$}
\centering
\footnotesize
 \label{tab3}
\begin{tabular}{c  rcccc  c rcccc}
		\hline
		    $\hat \mu^{1}(z)$   &  Bias & $\sqrt{\text{Var}}$   &   $\sqrt{\text{EVar}}$ & Cov90 & Cov95 & &  Bias & $\sqrt{\text{Var}}$   &  $\sqrt{\text{EVar}}$ & Cov90 & Cov95  \\
		\hline		
		       &  \multicolumn{11}{c}{  \multirow{2}{*}{
                            \cite{Fan-Hsu-Lieli-Zhang2019}'s AIPW  with  full sample} }   \\
                            &  \multicolumn{11}{c}{} \\
		  &  \multicolumn{5}{c}{  \multirow{1}{*}{(C4) cor PS, cor OR} }  & &   \multicolumn{5}{c}{  \multirow{1}{*}{(C5) cor PS, mis OR} }      \\
	    $\hat \mu^{1}(-0.4)$   &   -0.018 & 0.406 & 0.359 & 0.858 & 0.930 &  &  -0.057 & 0.210 & 0.190 & 0.837 & 0.899 \\
			 $\hat \mu^{1}(-0.2)$  & -0.036 & 0.355 & 0.349 & 0.912 & 0.943 &   & -0.037 & 0.191 & 0.184 & 0.861 & 0.917    \\
	    $\hat \mu^{1}(0.0)$  	&  -0.025 & 0.387 & 0.352 & 0.888 & 0.930 &  & -0.038 & 0.207 & 0.195 & 0.856 & 0.925 \\
			 $\hat \mu^{1}(0.2)$   &  -0.016 & 0.383 & 0.354 & 0.882 & 0.929 & &  -0.050 & 0.198 & 0.197 & 0.862 & 0.914  \\
			  $\hat \mu^{1}(0.4)$   &   -0.030 & 0.376 & 0.358 & 0.872 & 0.932 & & -0.050 & 0.218 & 0.209 & 0.828 & 0.881 \\
			   \\ \hline
			            &  \multicolumn{11}{c}{  \multirow{2}{*}{
                            \cite{Fan-Hsu-Lieli-Zhang2019} and \cite{Zimmert-Lechner2019}'s AIPW
                            with  four-fold cross-fitting} }   \\
                            &  \multicolumn{11}{c}{} \\
		  &  \multicolumn{5}{c}{  \multirow{1}{*}{(C4) cor PS, cor OR} }  & &   \multicolumn{5}{c} {  \multirow{1}{*}{(C5) cor PS, mis OR} }    \\
	    $\hat \mu^{1}(-0.4)$   &  -0.034 & 0.417 & 0.376 & 0.873 & 0.927 &  &  -0.047 & 0.296 & 0.300 & 0.826 & 0.899 \\
			 $\hat \mu^{1}(-0.2)$    &  -0.001 & 0.441 & 0.399 & 0.875 & 0.931 & & -0.034 & 0.286 & 0.264 & 0.867 & 0.926 \\
	    $\hat \mu^{1}(0.0)$  	&   -0.016 & 0.417 & 0.397 & 0.869 & 0.928 &  &  -0.027 & 0.266 & 0.255 & 0.857 & 0.917  \\
			 $\hat \mu^{1}(0.2)$   &  -0.032 & 0.388 & 0.377 & 0.892 & 0.942 & &-0.040 & 0.264 & 0.252 & 0.845 & 0.903 \\
			  $\hat \mu^{1}(0.4)$   &   0.027 & 0.443 & 0.432 & 0.882 & 0.940 && -0.009 & 0.266 & 0.242 & 0.848 & 0.908 \\
	\hline			
	\end{tabular}
\end{table}
\normalsize

			
\section{Application}  \label{sec-application}

Low birth weights can be closely related to {one's} health, education and performances in labor
market throughout life (\citep{Almond-Chay-Lee2005}; \citep{Almond-Currie2011}), and maternal smoking is regarded as the most important preventable cause for low birth weights  (\citep{Kramer1987}).  Many studies assessed the magnitude of maternal smoking on
birth weights by estimating its average treatment effect. See 
\citet{daVeiga-Wilder2008} and  \citet{Walker-Tekin-Wallace2009}.
In this empirical analysis, we are interested in  exploring the heterogeneity of effects of maternal smoking on infant birth weights across different subpopulations by analyzing a dataset available from the STATA website
{\color{blue} (http://www.stata-press.com/data/r13/cattaneo2.dta)}.  The same question  have been investigated in recent works of \citet{Abrevaya-Hsu-Lieli2015}, \citet{Lee-Okui-Whang2017}, \citet{Fan-Hsu-Lieli-Zhang2019} and \citet{Zimmert-Lechner2019}.

The proposed methods improve on these previous investigations by obtaining both doubly robust point estimators and model-assisted  confidence intervals,  which allows for one of the PS model or OR model misspecified and hence the results are more reliable.
In addition,  the proposed method handles discrete and continuous $Z$ in a unified manner, while the competing methods {are only focused on the case of continuous $Z$.}

 For ease of comparison, as done in \citet{Lee-Okui-Whang2017}, we restrict our sample to white and non-Hispanic mothers. This yields a data subset with size 3,754.  The outcome variable $Y$ is the infant birth weights measured in grams, the exposure variable $T$ is the status of maternal smoking: $T=1$ represents that the mother smokes and $T=0$ otherwise. The covariates $X$ include  the parents' socio-economic information and the mother's medical and health records,  which is summarized in Table \textcolor{blue}{S2} of the supplementary material. The subpopulations of interest are defined by covariates $Z$, which
   are taken to be \texttt{alcohol}, \texttt{deadkids},  \texttt{prenatal1}, \texttt{fbaby} or \texttt{mage}.
     We pick these variables because they may have an impact on  infant brith weights:  \texttt{alcohol} is a risk factor known for low birth weights; \texttt{deadkids} and \texttt{mage} may have influence on mother's  health;  \texttt{presental1} = 1 implies that the mother has a prenatal visit early. The first four variables are binary and the last one is continuous.

    As suggested in Section \ref{doubly-robust}, we set $f(X) = g(X) = (1, V^{T}, Z, V^{T} Z)^{T}$ for binary $Z$, where $V$ is all the covariates listed in Table \textcolor{blue}{S2} of the supplementary material excluding variable $Z$; the dimension of $f(X)$ is $p = 45$, excluding the constant. All variables in $f(X)$ are standardized to have sample mean 0 and sample variance 1.
     As done in simulation studies, the lasso tuning parameter $\lambda$ is selected by five-fold cross validation.
     Table \ref{tab5} summarizes the estimated causal effects of maternal smoking on infant birth weights conditional on different binary variable $Z$, where {CSTE} is the point estimator of $\tau(z)$, 
   SD and 95\% CI denote the estimated standard error and  95\% point-wise confidence interval 
   by using formula  (\ref{eq29}).   As shown in Table \ref{tab5},  there is a significant difference in causal effect point-estimators  between the first infant group (-177.40) and not first infant group (-322.97). And the alcohol consumed group (-290.14)  tends to have a lower infant birth weights than the group of not drinking (-259.06). The asymptotic variance of $\hat \tau(1)$ in alcohol consumed group is large, this is because only 93 subjects  in drinking group.  In addition,   the causal effects of maternal smoking on birth weights have no large difference among subgroups with different values of \texttt{deadkids} and \texttt{prenatal1}.

\begin{table}[H]
\caption{Effects of maternal smoking on infant birth weights for discrete $Z$ with proposed method}
\centering
\footnotesize
 \label{tab5}
\begin{tabular}{c  lll  c lll}
		\hline
		    $\hat \tau(z)$   & {CSTE} &  SD  &  95\% CI &   &     {CSTE} &  SD  &  95\% CI  \\
		\hline		
		  &  \multicolumn{3}{c}{ 	\multirow{2}{*}{$Z$ = alcohol} } & & \multicolumn{3}{c}{ 	\multirow{2}{*}{$Z$ = deadkids} }  \\
		  &  \multicolumn{7}{c}{} \\
		 $\hat \tau(0)$	& -259.06 & 33.36 & (-324.45, -193.67) &  & -270.56 & 30.27 & (-329.89, -211.23)  \\
	  	    $\hat \tau(1)$ &	-290.14 & 97.43 & (-481.10, -99.18)  & &  -251.28 & 54.56 & (-358.22, -144.34) \\
		    	  &  \multicolumn{3}{c}{ 	\multirow{2}{*}{$Z$ = prenatal1} } & & \multicolumn{3}{c}{ 	\multirow{2}{*}{$Z$ = fbaby} }  \\
		  &  \multicolumn{7}{c}{} \\
		 $\hat \tau(0)$	&  -275.59 & 47.90 & (-369.47, -181.71) & & -322.97 & 41.72 & (-404.74, -241.20) \\
	  	    $\hat \tau(1)$	& -263.88 & 35.63 & (-333.71, -194.05) & &  -177.40 & 49.03 & (-273.50, -81.30) \\
		\hline			
	\end{tabular}
	\begin{flushleft}
	  Note: {CSTE} is the point estimator of $\tau(z) = E(Y^{1} - Y^{0}| Z= z)$, SD and 95\% CI denote the estimated standard error and  95\% point-wise confidence interval of $\hat \tau(z)$, respectively. 
	\end{flushleft}
\end{table}
\normalsize

{Given that the existing methods in Section \ref{sec-existing-estimators} {do not deal with discrete $Z$, we compare the proposed method with a naturally alternative approach, which also employs the same basis functions $\Phi(z)$ and regressors $f(X) = g(X) = (1, V^{T}, Z)^{T}$, while using regularized maximum likelihood estimators $\hat \pi_{RML}$ and $\hat m_{t, RML}$ for $t=0, 1$.} The corresponding results are similar to those in Table \ref{tab5} and are presented in Table \textcolor{blue}{S3} of the supplementary material.}

\begin{figure}[h!]
 \centering
 \includegraphics[scale = 0.6]{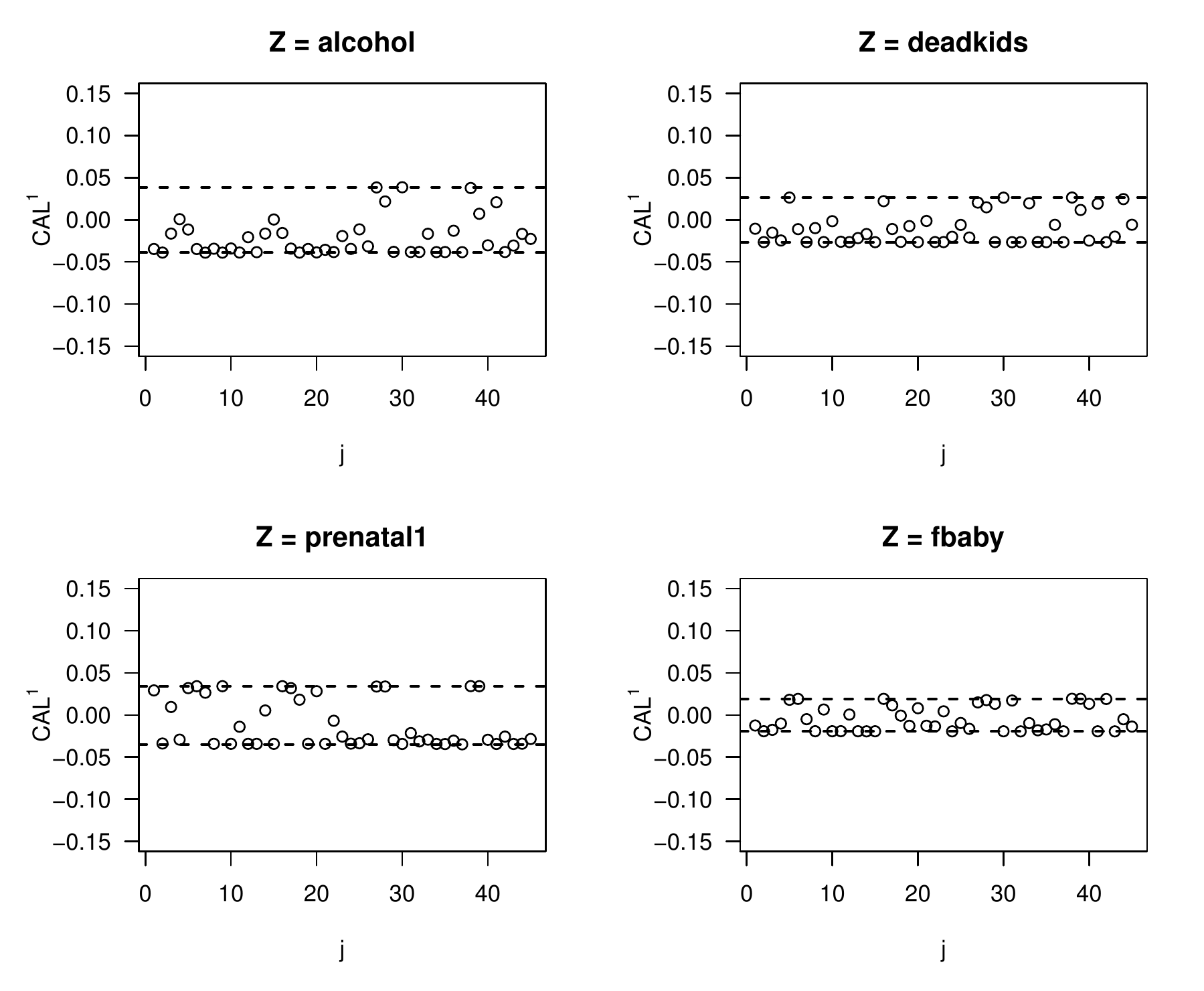}
 \caption{Standardized calibration differences $\text{CAL}^{1}(\hat \pi_{RCAL}; f_{j})$ plotted against index $j$ for $\hat \pi_{RCAL}$ with $\lambda$ selected by 5-fold cross-validation} 
 \label{fig1}
\end{figure}

\begin{figure}[h!]
 \centering
 \includegraphics[scale = 0.6]{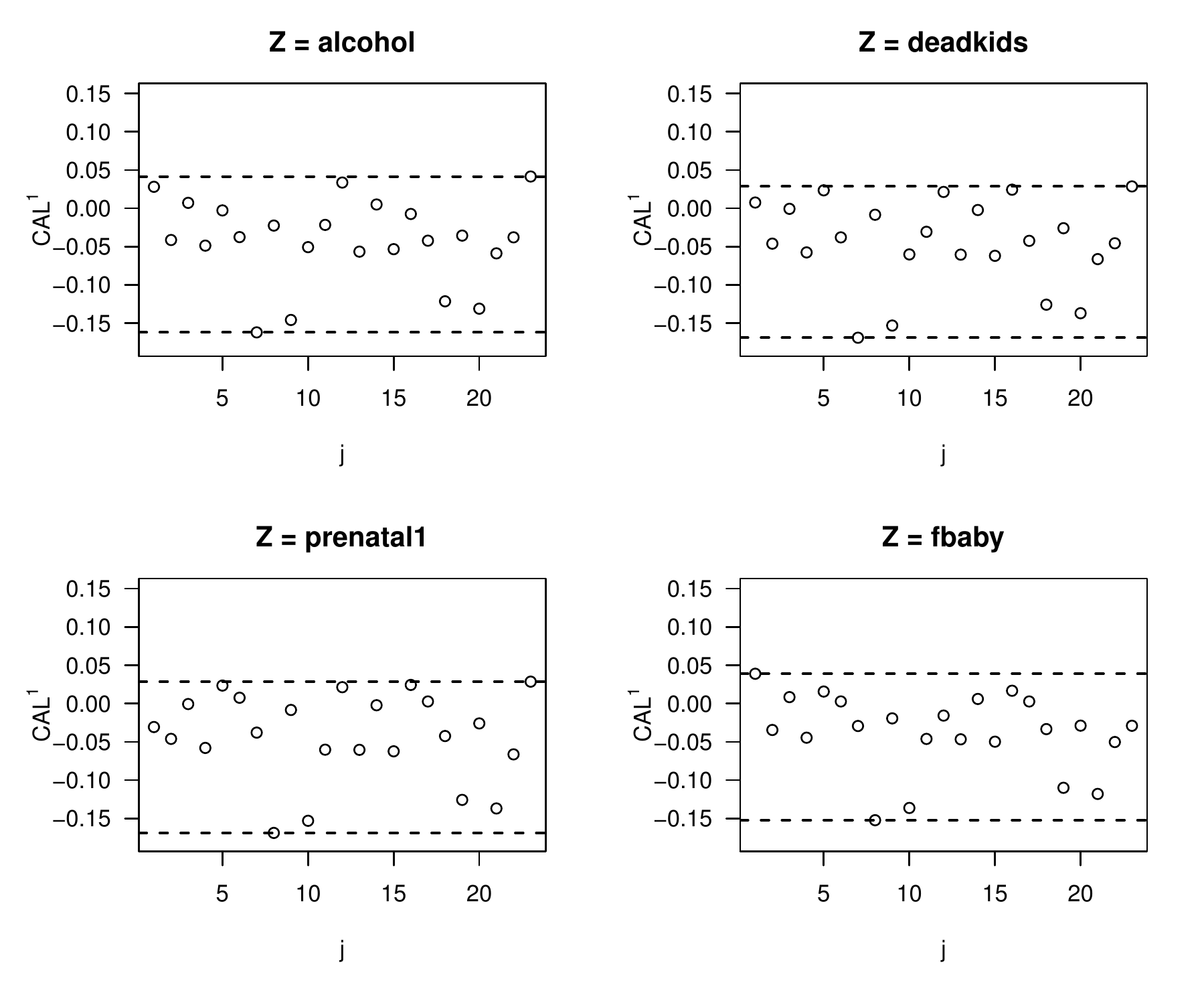}
 \caption{Standardized calibration differences $\text{CAL}^{1}({\hat \pi_{RML}}; f_{j})$ plotted against index $j$ for {$\hat \pi_{RML}$} with $\lambda$ selected by 5-fold cross-validation.} 
 \label{fig2}
\end{figure}

We conduct balance checking by using a propensity score estimate $\hat \pi_{RCAL}$ in treated sample to assess the reliability of results in Table \ref{tab5}.  Specifically, for a function $h(X)$,  we use the standardized calibration difference {(\citep{Tan-BKA2020})}  
		$$\text{CAL}^{1}( \hat \pi; h) =\Big [  \frac{\tilde E \{ T h(X)   \hat \pi^{-1}(X) \}   }{ \tilde E \{ T  \hat \pi^{-1}(X) \} }    -  \tilde E\{h(X)\}  \Big  ] \biggl / \tilde V^{1/2}\{ h(X)\}$$
to measure the effect of calibration, where $\tilde E(\cdot)$ and $\tilde V(\cdot)$ denote the sample mean and variance.
 Figure \ref{fig1} and \ref{fig2} display the values of  standardized calibration difference of all covariates for the four choices of binary $Z$, obtained from  $\hat \pi_{RCAL}$ and $\hat \pi_{RML}$.  It can be seen that the values of $\text{CAL}^{1}(\hat \pi_{RCAL}; f_{j})$ tend be less variable than those of $\text{CAL}^{1}(\hat \pi_{RML}; f_{j})$ for different $Z$, which implies that $\hat \pi_{RCAL}$
	can  balance covariates better than $\hat \pi_{RML}$ and thus the associated results are more reliable. 

         To estimate a {CSTE} curve when $Z$ is mage, we apply the propose method  using cubic spline to approximate $\tau(z)$ and find the optimal number of knots by using grid search with Akaike information criterion (AIC, \citep{Akaike1974})  and Bayesian information criterion (BIC, \citep{Schwarz2005}). Specifically,  we first fix the number of knots is 10, that is, $f(X) = (1, V^{T}, \Phi(Z)^{T})^{T}$, $g(X) = ( 1, V^{T}, \Phi(Z)^{T},  (V\otimes \Phi(Z))^{T},  (\Phi(Z) \otimes \Phi(Z))^{T} )^{T}$ with $\Phi(Z)$ being cubic spline basis functions with 10 knots. Then we use $f(X)$ and $g(X)$ to estimate propensity score and outcome regression functions. Finally, we conduct least squares by regressing $\varphi( Y, T, X; \hat m_{1}, \hat \pi)  - \varphi( Y, 1 - T, X; \hat m_{0}, 1 - \hat \pi_{0})$ on $\tilde \Phi(Z)$ to get the values of AIC and BIC, where  $\tilde \Phi(Z)$ is cubic spline basis functions with number of knots ranging from 1 to 10.   As can be seen from Table \ref{tab6}, the optimal choice of number of knots is 4 for both AIC and BIC.
\begin{table}[H]
\caption{Values of AIC and BIC under different number of knots}
\centering
\footnotesize
 \label{tab6}
\begin{tabular}{ccc ccc}
		\hline
		    $K$  &  AIC   & BIC    &  $K$  &  AIC   & BIC    \\
		\hline		
1 & 67290.35 & 67327.68 &   6 & 67293.03 & 67361.45 \\
    2 & 67287.36 & 67330.90 &   7 & 67299.65 & 67374.29 \\
    3 & 67292.36 & 67342.12 &   8 & 67662.25 & 67743.11 \\
    4 & 66600.08 & 66656.06 &   9 & 66609.08 & 66696.16 \\
    5 & 67290.88 & 67353.08 &  10 & 67304.51 & 67397.82 \\
		\hline			
	\end{tabular}
\begin{center} Note: $K$ is the number of knots for cubic spline. \end{center}
\end{table}
\normalsize

 \begin{figure}[h!]
 \centering
 \includegraphics[scale = 0.7]{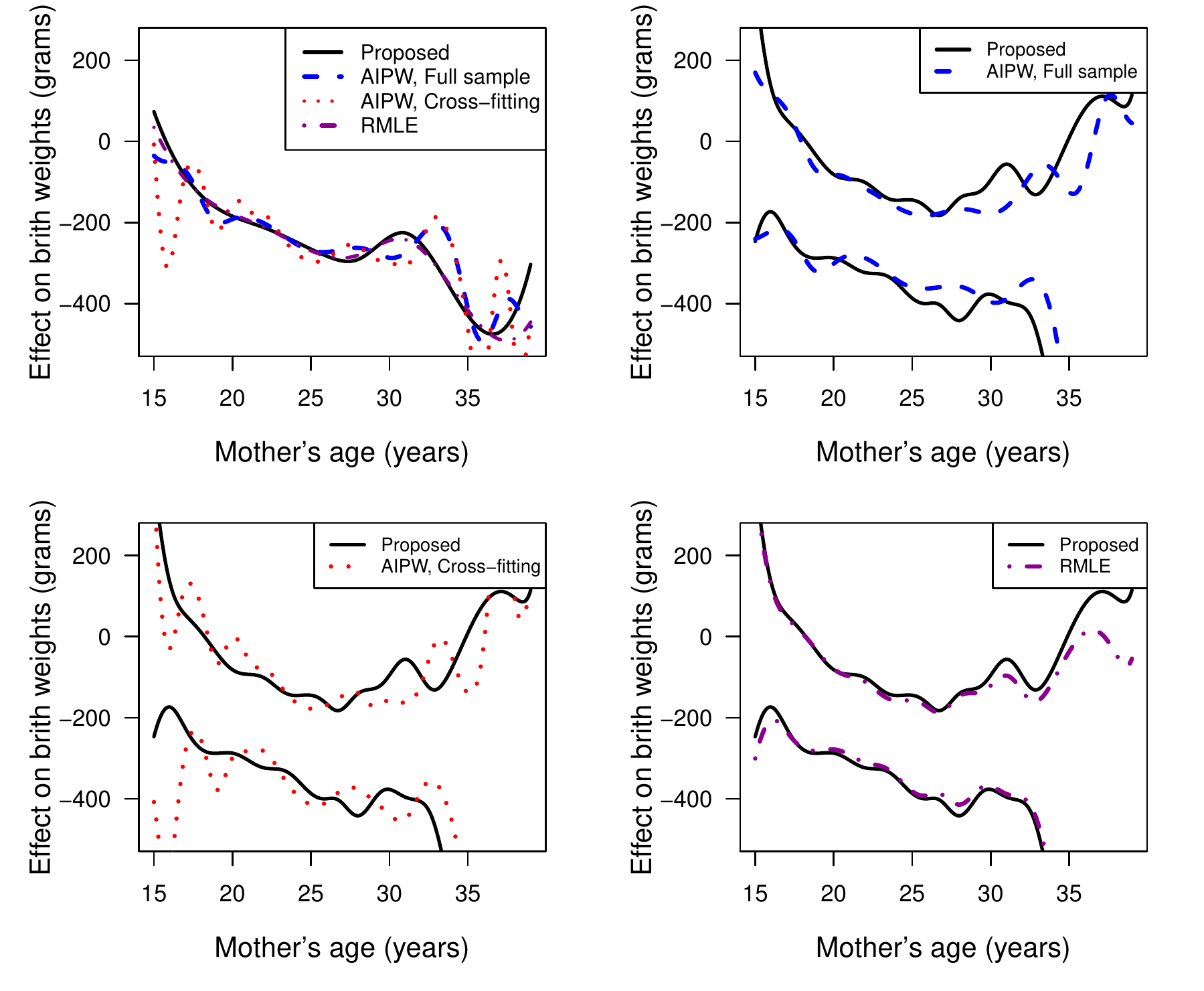}
 \caption{upper left: Estimated {CSTE} curves; upper right: 95\% CI for Full sample; lower left: 95\% CI for Cross-Fitting; lower right: 95\% CI for RMLE.}
 \label{fig4}
\end{figure}

\begin{figure}[h!]
 \centering
 \includegraphics[scale = 0.6]{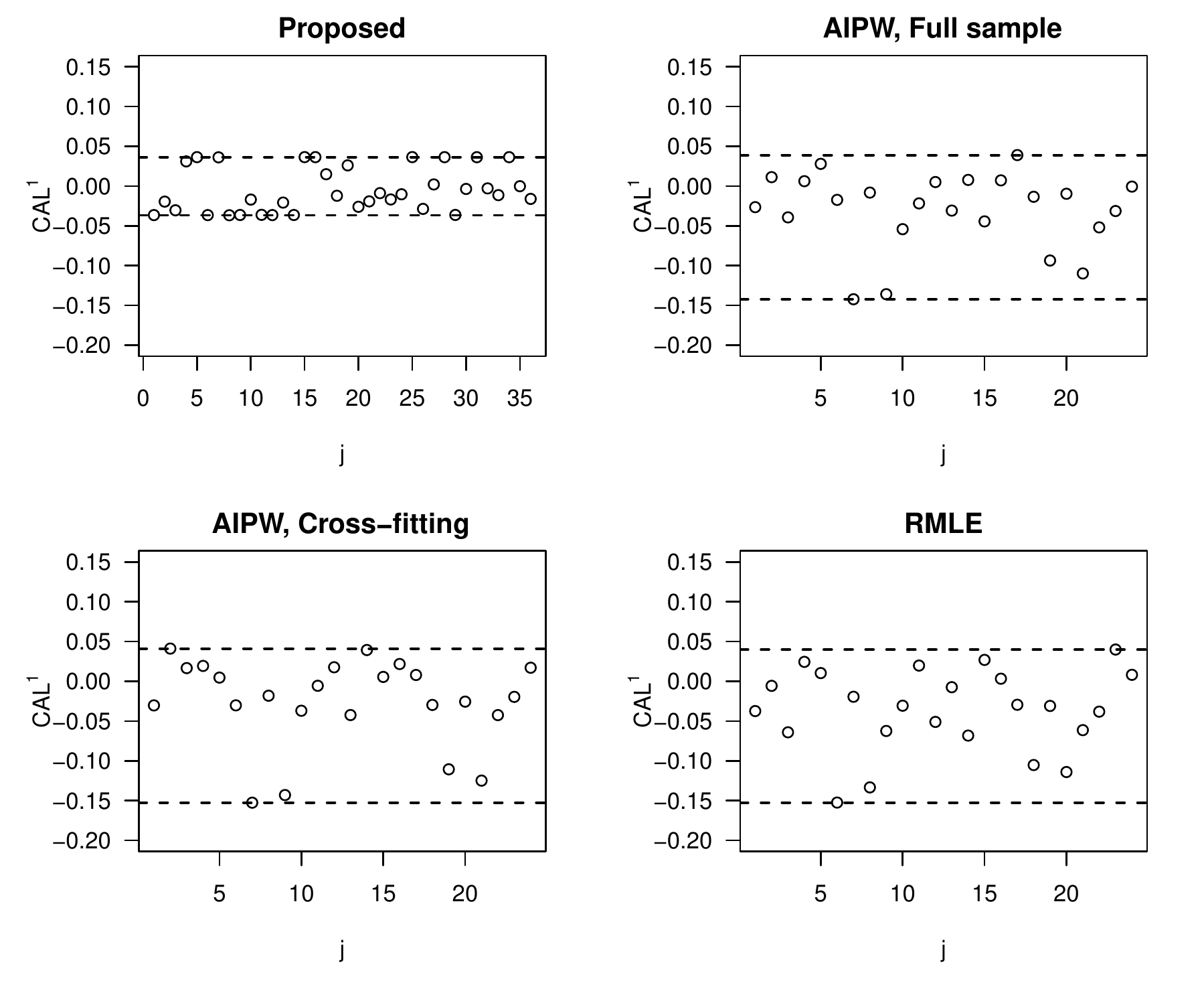}
 \caption{Standardized calibration differences of four competing methods (Proposed, AIPW with full sample, AIPW with cross-fitting, RMLE) for continuous $Z$.}
 \label{fig5}
\end{figure}

Figure \ref{fig4} displays the resulting estimates of CATE curves, where
Figure \ref{fig4}(a) presents the point estimates,  Figure \ref{fig4}(b) and \ref{fig4}(c) show the associated 95\% point-wise confidence intervals.
  As done in the simulation with continuous $Z$ and the application with discrete $Z$, we include the competing AIPW estimates of  \citet{Fan-Hsu-Lieli-Zhang2019} and \citet{Zimmert-Lechner2019} and
  {regularized maximum likelihood estimate (RMLE)} for comparison.    	
	As shown in Figure \ref{fig4}, all methods produce similar trends in point estimates and confidence intervals, although they differ in finer scales.
The AIPW, cross-fitting method appears to yield large variations at the two ends of the age interval.
	{Figure \ref{fig5} depicts the standardized calibration differences for the four methods, which also shows that $\hat \pi_{RCAL}$ {can better balance all covariates and hence lead to 
more reliable results than the other methods}.}
	Overall, all three methods demonstrate that maternal smoking has a negative effect on birth weights, and  the negative effect becomes stronger as age increases before 35. The estimated CATE curve has a drastic change and large variance with age ranging from 35 to 40, which may be induced by a small sample size in this age range.
	 This is consistent with the findings in  previous studies.


%

\section{Discussion}  \label{sec-discussion}
This article develops new methods  to obtain both doubly robust point estimators and model-assisted confidence intervals for conditional average treatment effects in high-dimensional settings.
 In addition, with a linear OR model and discrete $Z$,   the confidence intervals are also doubly robust.
  Theoretical properties are established for the proposed methods with different data types of outcome $Y$ and covariates $Z$, and the corresponding variances can be estimated by a sandwich method.

{Further work is desired to extend our method and theory by relaxing the parametric structural model (\ref{eq16}) to be nonparametric subject to smoothness conditions, while
allowing the basis functions $\Phi(z)$ to be data-adaptively chosen, instead of pre-specified.}
 Another interesting question is that whether  doubly robust confidence intervals can be derived for continuous $Z$.
  To deal with this question, {a possible approach} is to discretize $Z$.
For example, for two knots  $t_{1} < t_{2}$, we can discretize $Z$ as $(Z_{1}, Z_{2}) = (I\{t_{1} < Z \leq t_{2} \}, I\{ Z > t_{2} \} )$ or $(Z_{1}, Z_{2}) = (I\{ Z >  t_{1}  \}, I\{ Z > t_{2} \} )$.
With either choice of $(Z_{1}, Z_{2})$, the proposed method using $f(X) =  ( 1, V^{T}, Z_{1},  Z_{2},  V^{T}Z_{1},  V^{T} Z_{2}  )^{T}$ achieves desired doubly robust confidence intervals in the
discretized model
	  \[     \mu^{1}(Z)  =  E(Y^{1} | Z ) =  \beta_{0} + (Z_{1} ,  Z_{2} ) \beta_{1}.   \]
  The method can be easily extended to multiple knots, corresponding to piecewise constant model for $\mu^{1}(z)$.
  Then various theoretical questions need to be investigated.    For example, it is interesting to study convergence and whether doubly confidence intervals can be achieved, depending on the number of knots used.

Another extension is to consider the case that $Z$ is composed of multiple continuous variables.
 A possible strategy is to postulate an additive model (\citep{Hastie-Tibshirani1990})
	  		\begin{equation*}
				\mu^{1}(Z)   =  \beta_{0} +  \sum_{k=1}^{m}  g_{k}(Z_{k}),
		\end{equation*}
where  $g_{k}$ is an unknown function of the $j$-th covariate. Alternatively, we may consider a single index model (\citep{Guo-Zhou-Ma2021})
	\begin{equation*}
			 	\mu^{1}(Z)   =   g(Z^{T} \beta).
	\end{equation*}
It is interesting to study how to incorporate such strategies in future research.

 \bibliography{ref}
 \bibliographystyle{agsm}

\end{spacing}
\end{document}